\documentclass[twocolumn,twocolappendix]{aastex63}
\usepackage{natbib,aas_macros,amsmath}
\usepackage{multirow,color}
\usepackage{natbib}
\usepackage{here}
\usepackage{soul}
\usepackage{lineno}

\newcommand{\oiii}{[O\,{\sc iii}]}

\newcommand{\cii}{[C\,{\sc ii}]}
\newcommand{\nii}{[N\,{\sc ii}]}

\newcommand{\hst}{{\it HST}}
\newcommand{\jwst}{{\it JWST}}
\accepted{in ApJ}
\shorttitle{
ALMA FIR View of $z\sim$11--17 candidates
}
\shortauthors{Fujimoto et al.}

\begin{document}
\title{
ALMA FIR View of Ultra High-redshift Galaxy Candidates at $z\sim$ 11--17: \\
Blue Monsters or Low-$z$ Red Interlopers$?$
}


\correspondingauthor{Seiji Fujimoto}
\email{fujimoto@utexas.edu}
\author[0000-0001-7201-5066]{Seiji Fujimoto}\altaffiliation{Hubble Fellow}
\affiliation{Department of Astronomy, The University of Texas at Austin, Austin, TX, USA}
\affiliation{
Cosmic Dawn Center (DAWN), Jagtvej 128, DK2200 Copenhagen N, Denmark
}
\affiliation{
Niels Bohr Institute, University of Copenhagen, Lyngbyvej 2, DK2100 Copenhagen \O, Denmark
}

\author[0000-0001-8519-1130]{Steven L. Finkelstein}
\affiliation{Department of Astronomy, The University of Texas at Austin, Austin, TX, USA}

\author[0000-0002-4193-2539]{Denis Burgarella}
\affiliation{Aix Marseille Univ, CNRS, CNES, LAM Marseille, France}

\author[0000-0001-6647-3861]{Chris L. Carilli}
\affiliation{National Radio Astronomy Observatory, P.O. Box O, Socorro, NM, 87801, USA}


\author[0000-0003-3441-903X]{V\'eronique Buat}
\affiliation{Aix Marseille Univ, CNRS, CNES, LAM Marseille, France}

\author[0000-0002-0930-6466]{Caitlin M. Casey}
\affiliation{Department of Astronomy, The University of Texas at Austin, Austin, TX, USA}

\author[0000-0003-0541-2891]{Laure Ciesla}
\affiliation{Aix Marseille Univ, CNRS, CNES, LAM Marseille, France}

\author[0000-0002-8224-4505]{Sandro Tacchella}
\affiliation{Kavli Institute for Cosmology, University of Cambridge, Madingley Road, Cambridge, CB3 0HA, UK}
\affiliation{Cavendish Laboratory, University of Cambridge, 19 JJ Thomson Avenue, Cambridge, CB3 0HE, UK}

\author[0000-0002-7051-1100]{Jorge A. Zavala}
\affiliation{National Astronomical Observatory of Japan, 2-21-1 Osawa, Mitaka, Tokyo 181-8588, Japan}


\author[0000-0003-2680-005X]{Gabriel Brammer}
\affiliation{
Cosmic Dawn Center (DAWN), Jagtvej 128, DK2200 Copenhagen N, Denmark
}
\affiliation{
Niels Bohr Institute, University of Copenhagen, Lyngbyvej 2, DK2100 Copenhagen \O, Denmark
}


\author[0000-0001-6477-4011]{Yoshinobu Fudamoto}
\affiliation{
Waseda Research Institute for Science and Engineering, Faculty of Science and Engineering, Waseda University, 3-4-1 Okubo, Shinjuku, Tokyo 169-8555, Japan
}
\affiliation{
National Astronomical Observatory of Japan, 2-21-1, Osawa, Mitaka, Tokyo, Japan
}

\author[0000-0003-3441-903X]{Masami Ouchi}
\affiliation{Institute for Cosmic Ray Research, The University of Tokyo, 5-1-5 Kashiwanoha, Kashiwa, Chiba 277-8582, Japan}
\affiliation{National Astronomical Observatory of Japan, 2-21-1 Osawa, Mitaka, Tokyo 181-8588, Japan}
\affiliation{
Kavli Institute for the Physics and Mathematics of the Universe (WPI), University of Tokyo, Kashiwa, Chiba 277-8583, Japan}

\author[0000-0001-6477-4011]{Francesco Valentino}
\affiliation{
Cosmic Dawn Center (DAWN), Jagtvej 128, DK2200 Copenhagen N, Denmark
}
\affiliation{
Niels Bohr Institute, University of Copenhagen, Lyngbyvej 2, DK2100 Copenhagen \O, Denmark
}
\affiliation{
European Southern Observatory, Karl-Schwarzschild-Str. 2, D-85748, Garching, Germany
}

\author[0000-0003-1371-6019]{M. C. Cooper}
\affiliation{Department of Physics \& Astronomy, University of California, Irvine, 4129 Reines Hall, Irvine, CA 92697, USA}

\author[0000-0001-5414-5131]{Mark Dickinson}
\affiliation{NSF's National Optical-Infrared Astronomy Research Laboratory, 950 N. Cherry Ave., Tucson, AZ 85719, USA}

\author[0000-0002-3560-8599]{Maximilien Franco}
\affiliation{Department of Astronomy, The University of Texas at Austin, Austin, TX, USA}

\author[0000-0002-7831-8751]{Mauro Giavalisco}
\affiliation{University of Massachusetts Amherst, 710 North Pleasant Street, Amherst, MA 01003-9305, USA}

\author[0000-0001-6251-4988]{Taylor A. Hutchison}
\altaffiliation{NASA Postdoctoral Fellow}
\affiliation{Astrophysics Science Division, NASA Goddard Space Flight Center, 8800 Greenbelt Rd, Greenbelt, MD 20771, USA}

\author[0000-0001-9187-3605]{Jeyhan S. Kartaltepe}
\affiliation{Laboratory for Multiwavelength Astrophysics, School of Physics and Astronomy, Rochester Institute of Technology, 84 Lomb Memorial Drive, Rochester, NY 14623, USA}

\author[0000-0002-6610-2048]{Anton M. Koekemoer}
\affiliation{Space Telescope Science Institute, 3700 San Martin Drive, Baltimore, MD 21218, USA}

\author[0000-0001-5780-1886]{Takashi Kojima}
\affiliation{Institute for Cosmic Ray Research, The University of Tokyo, 5-1-5 Kashiwanoha, Kashiwa, Chiba 277-8582, Japan}

\author[0000-0003-2366-8858]{Rebecca L. Larson}
\affiliation{NSF Graduate Fellow}
\affiliation{Department of Astronomy, The University of Texas at Austin, Austin, TX, USA}

\author[0000-0001-7089-7325]{E.~J. Murphy}
\affiliation{National Radio Astronomy Observatory, 520 Edgemont Road, Charlottesville, VA 22903, USA} 

\author[0000-0001-7503-8482]{Casey Papovich}
\affiliation{Department of Physics and Astronomy, Texas A\&M University, College Station, TX, 77843-4242 USA}
\affiliation{George P.\ and Cynthia Woods Mitchell Institute for Fundamental Physics and Astronomy, Texas A\&M University, College Station, TX, 77843-4242 USA}

\author[0000-0003-4528-5639]{Pablo G. P\'erez-Gonz\'alez}
\affiliation{Centro de Astrobiolog\'{\i}a (CAB/CSIC-INTA), Ctra. de Ajalvir km 4, Torrej\'on de Ardoz, E-28850, Madrid, Spain}

\author[0000-0002-6748-6821]{Rachel S.~Somerville}
\affiliation{Center for Computational Astrophysics, Flatiron Institute, 162 5th Avenue, New York, NY 10010, USA}

\author[0000-0001-9163-0064]{Ilsang Yoon}
\affiliation{National Radio Astronomy Observatory, 520 Edgemont Road, Charlottesville, VA 22903, USA}

\author[0000-0003-3903-6935]{Stephen M.~Wilkins} %
\affiliation{Astronomy Centre, University of Sussex, Falmer, Brighton BN1 9QH, UK}
\affiliation{Institute of Space Sciences and Astronomy, University of Malta, Msida MSD 2080, Malta}


\author[0000-0002-7821-8873]{Hollis Akins}
\affiliation{Department of Astronomy, The University of Texas at Austin, Austin, TX, USA}

\author[0000-0001-5758-1000]{Ricardo O. Amor\'{i}n}
\affiliation{Instituto de Investigaci\'{o}n Multidisciplinar en Ciencia y Tecnolog\'{i}a, Universidad de La Serena, Raul Bitr\'{a}n 1305, La Serena 2204000, Chile}
\affiliation{Departamento de Astronom\'{i}a, Universidad de La Serena, Av. Juan Cisternas 1200 Norte, La Serena 1720236, Chile}

\author[0000-0002-7959-8783]{Pablo Arrabal Haro}
\affiliation{NSF's National Optical-Infrared Astronomy Research Laboratory, 950 N. Cherry Ave., Tucson, AZ 85719, USA}

\author[0000-0002-9921-9218]{Micaela B. Bagley}
\affiliation{Department of Astronomy, The University of Texas at Austin, Austin, TX, USA}

\author[0000-0003-4922-0613]{Katherine Chworowsky}
\affiliation{Department of Astronomy, The University of Texas at Austin, Austin, TX, USA}

\author[0000-0001-7151-009X]{Nikko J. Cleri}
\affiliation{Department of Physics and Astronomy, Texas A\&M University, College Station, TX, 77843-4242 USA}
\affiliation{George P.\ and Cynthia Woods Mitchell Institute for Fundamental Physics and Astronomy, Texas A\&M University, College Station, TX, 77843-4242 USA}

\author[0000-0003-3881-1397]{Olivia R. Cooper}
\affiliation{Department of Astronomy, The University of Texas at Austin, 2515 Speedway Boulevard Stop C1400, Austin, TX 78712, USA}

\author[0000-0001-6820-0015]{Luca Costantin}
\affiliation{Centro de Astrobiolog\'ia (CSIC-INTA), Ctra de Ajalvir km 4, Torrej\'on de Ardoz, 28850, Madrid, Spain}

\author[0000-0002-3331-9590]{Emanuele Daddi}
\affiliation{Universit\'e Paris-Saclay, Universit\'e Paris Cit\'e, CEA, CNRS, AIM, 91191, Gif-sur-Yvette, France}

\author[0000-0001-7113-2738]{Henry C. Ferguson}
\affiliation{Space Telescope Science Institute, Baltimore, MD, USA}

\author[0000-0001-9440-8872]{Norman A. Grogin}
\affiliation{Space Telescope Science Institute, 3700 San Martin Drive, Baltimore, MD 21218, USA}

\author[0000-0002-2640-5917]{E.~F. Jim\'enez-Andrade}
\affiliation{Instituto de Radioastronomía y Astrofísica, UNAM Campus Morelia, Apartado postal 3-72, 58090 Morelia, Michoacán, México}

\author[0000-0002-0000-2394]{St{\'e}phanie Juneau}
\affiliation{NSF's NOIRLab, 950 N. Cherry Ave., Tucson, AZ 85719, USA}

\author[0000-0002-5537-8110]{Allison Kirkpatrick}
\affiliation{Department of Physics and Astronomy, University of Kansas, Lawrence, KS 66045, USA}

\author[0000-0002-8360-3880]{Dale D. Kocevski}
\affiliation{Department of Physics and Astronomy, Colby College, Waterville, ME 04901, USA}

\author[0000-0002-9466-2763]{Aur{\'e}lien Le Bail}
\affil{Universit{\'e} Paris-Saclay, Université Paris Cit{\'e}, CEA, CNRS, AIM, 91191, Gif-sur-Yvette, France}

\author[0000-0002-7530-8857]{Arianna Long}\altaffiliation{Hubble Fellow}
\affiliation{Department of Astronomy, The University of Texas at Austin, Austin, TX, USA}

\author[0000-0003-1581-7825]{Ray A. Lucas}
\affiliation{Space Telescope Science Institute, 3700 San Martin Drive, Baltimore, MD 21218, USA}

\author[0000-0002-6777-6490]{Benjamin Magnelli}
\affiliation{Universit\'e Paris-Saclay, Universit\'e Paris Cit\'e, CEA, CNRS, AIM, 91191, Gif-sur-Yvette, France}

\author[0000-0002-6149-8178]{Jed McKinney}
\affiliation{Department of Astronomy, The University of Texas at Austin, Austin, TX, USA}

\author[0000-0002-8018-3219]{Caitlin Rose}
\affil{Laboratory for Multiwavelength Astrophysics, School of Physics and Astronomy, Rochester Institute of Technology, 84 Lomb Memorial Drive, Rochester, NY 14623, USA}

\author[0000-0001-7755-4755]{Lise-Marie Seill\'e}
\affiliation{Aix Marseille Univ, CNRS, CNES, LAM Marseille, France}

\author[0000-0002-6386-7299]{Raymond C. Simons}
\affiliation{Space Telescope Science Institute, 3700 San Martin Drive, Baltimore, MD 21218, USA}

\author[0000-0001-6065-7483]{Benjamin J. Weiner}
\affiliation{MMT/Steward Observatory, University of Arizona, 933 N. Cherry St, Tucson, AZ 85721, USA}

\author[0000-0003-3466-035X]{{L. Y. Aaron} {Yung}}
\altaffiliation{NASA Postdoctoral Fellow}
\affiliation{Astrophysics Science Division, NASA Goddard Space Flight Center, 8800 Greenbelt Rd, Greenbelt, MD 20771, USA}

\def\apj{ApJ}%
\def\apjl{ApJ}%
\def\apjs{ApJS}%

\def\rme{\rm e}
\def\rmstar{\rm star}
\def\rmFIR{\rm FIR}
\def\itHubble{\it Hubble}
\def\rmyr{\rm yr}
\def\targa{CEERS-93316}%
\def\targb{S5-z17-1}


\begin{abstract}
We present ALMA Band~7 observations of a remarkably bright galaxy candidate at $z_{\rm phot}$=$16.7^{+1.9}_{-0.3}$ ($M_{\rm UV}$=$-21.6$), S5-z17-1, identified in \jwst\ Early Release Observation data of Stephen's Quintet.
We do not detect the dust continuum at 866~$\mu$m, 
ruling out the possibility that \targb\ is a low-$z$ dusty starburst with a  star-formation rate of $\gtrsim 30$~$M_{\odot}$~yr$^{-1}$. 
We detect a 5.1$\sigma$ line feature at $338.726\pm0.007$~GHz exactly coinciding with the \jwst\ source position, 
with a 2\% likelihood of the signal being spurious.
The most likely line identification would be 
\oiii52$\mu$m at $z=16.01$ or \cii158$\mu$m at $z=4.61$, whose line luminosities do not violate the non-detection of the dust continuum in both cases.
Together with three other $z\gtrsim$ 11--13 candidate galaxies recently observed with ALMA, we conduct a joint ALMA and \jwst\ spectral energy distribution (SED) analysis and find that the high-$z$ solution at $z\sim$11--17 is favored in every candidate 
as a very blue (UV continuum slope of $\simeq-2.3$) and luminous ($M_{\rm UV}\simeq[-$24:$-21]$) system. 
Still, we find in several candidates that reasonable SED fits ($\Delta$ $\chi^{2}\lesssim4$) are reproduced by type-II quasar and/or quiescent galaxy templates with strong emission lines at $z\sim3$--5, 
where such populations predicted from their luminosity functions and EW(\oiii+H$\beta$) distributions are abundant in survey volumes used for the identification of the $z\sim$11--17 candidates.
While these recent ALMA observation results have strengthened the likelihood of the high-$z$ solutions, lower-$z$ possibilities are not completely ruled out in several of the $z\sim$11--17 candidates, indicating the need to consider the relative surface densities of the lower-$z$ contaminants in the ultra high-$z$ galaxy search.
\end{abstract}
\keywords{ galaxies: formation --- galaxies: evolution --- galaxies: high-redshift }

\section{Introduction}
\label{sec:intro} 

One of the major goals in modern astronomy is to understand when and how the first stars, black holes, and galaxies emerged in the universe. 
Despite the effort of exploring high redshifts at $z>10$ -- the first few hundred million years in our history of the universe -- only a single galaxy has been spectroscopically confirmed (GNz11 at $z\simeq11$; \citealt{oesch2016, jiang2021}). 
Because characterizing this first of stars and galaxies would bring a unique knowledge on the very first stellar populations and their impact on the early phases of galaxy evolution, and on the reionization, pushing this redshift frontier to the brink of the Big Bang and revealing the objects in the very first generations is a key driver of observational cosmology. 

From its first few weeks of science operations and months by now, \textit{James Webb Space Telescope} (\textit{JWST}) has sparked a revolution of the effort to discover and study galaxies at very early cosmic epochs. 
Three early \textit{JWST} observing programs have been carried out, and the data has been immediately public:  Early Release Observations (ERO, \citealt{pontoppidan2022}, PID 2736) for the gravitational lens galaxy cluster SMACS~J0723.3-7327 and Stephan's quintet field, and two Director's Discretionary Early Release Science (DD-ERS) programs: GLASS-JWST (PID 1324) and CEERS (PID 1345). All three programs include NIRCam imaging through multiple filters from 1 to 5 $\mu$m, suitable for identification of candidates for very high redshift objects using photometric redshifts and/or multi-color selection criteria \citep[e.g.,][]{adams2022, atek2022, castellano2022, donnan2022, finkelstein2022b, harikane2023, labbe2022, morishita2022, naidu2022, yan2022, bouwens2022c}. 
Discounted initial zeropoint calibration issues, their number and brightness are surprising and considerably exceeds most pre-\textit{JWST} predictions \citep[e.g.,][]{ferrara2022, finkelstein2023, mason2022}. 
These results indicate either the early universe was more prolific at forming galaxies than modern simulations predict with a potential strong implication on galaxy formation models (e.g., Finkelstein et al. 2022c, in prep), or there is significant foreground contamination in these early \textit{JWST} high-$z$ samples.

In this context, two of the most unique, highest-$z$ candidates are \targa\ and \targb\ identified in the CEERS and Stephan's Quintet fields, respectively \citep{donnan2022, harikane2023}. These candidates exhibit a clear ``dropout'' color signature and blue continuum slopes in NIRCam filters, interpreted as the redshifted Lyman-alpha break at $z\simeq$ 17 in both sources. 
These candidates are securely detected in the NIRCam filters at $>10\sigma$ levels with remarkably bright ultra-violet (UV) magnitudes of 26.3~mag and 26.6~mag (AB), corresponding to the absolute UV magnitudes of $\sim-22$ at $z=17$. 
In addition to the DSFG population, 
\cite{naidu2022b} argue that the similar NIRCam photometry is also reproduced by the active galactic nuclei (AGN) in quiescent galaxies at $z\sim5$, with an additional environmental evidence: 
all three of the galaxy's nearest neighbors at $< 2{\farcs}5$ have photometric redshifts of $z\sim5$; and the object could lie in a $z\sim5$ galaxy overdensity that is $\sim$5$\times$ overdense compared to the field.

Recent ALMA observations have detected millimeter emission from a significant population of ``$H$-dropout'' galaxies, undetected in \textit{HST} WFC3-IR imaging, with the dropout feature even by $>3$~mag between the \textit{HST}/F160W and \textit{Spitzer}/IRAC ch1 \citep[e.g.,][]{twang2019}. These galaxies are most likely massive DSFGs at $z\sim$3--5 (e.g., \citealt{fujimoto2016, franco2018, yamaguchi2019,  twang2019, williams2019, sun2021, barrufet2022, perez-gonzalez2022, rodighiero2022}). 
Moreover, these optical and near-infrared (NIR) faint DSFGs have been routinely identified in a serendipitous manner, originally targeting nearby massive galaxies (e.g., \citealt{romano2020, fudamoto2021, fujimoto2022}). This implies that the presence of the optical--NIR faint DSFGs traces the massive dark matter halos in the early universe (e.g., \citealt{twang2019, zhou2020}).
Therefore, the tentative SCUBA2 detection and the potential overdensity environment are in line with the properties of the $z\sim$3--5 DSFGs recently identified in the \textit{H}-dropout objects. 
Before concluding that \targa\ and \targb\ are remarkably bright $z\sim17$ galaxies, it is essential to rule out or confirm the lower-$z$ solution via further observations. 

In this paper, 
we present ALMA Band~7 DDT follow-up for \targb\ that is one of these remarkably UV bright $z\sim17$ candidates discovered in \jwst\ Early Release Observation data of Stephan's Quintet. 
This is the first FIR characterization of either of these $z \sim$ 17 candidates with ALMA\footnote{\targa\ is too far north to be accessible by ALMA and has been observed in NOEMA DDT (\#D22AC, PI: S. Fujimoto; see \citealt{arrabal-halo2023a})}, setting the benchmark to understand and interpret similarly high-$z$ candidates identified in the future \jwst\ observations. 
The structure of this paper is as follows. 
In Section 2, we describe the observations and the data reduction of both \jwst\ and ALMA. 
Section 3 outlines the methods and presents the results of the continuum flux measurements, a search for any emission line, and a full spectral energy distribution (SED) analysis, including other three galaxy candidates at $z\sim11$--13 recently observed with ALMA (GHZ1/GLz11, GHZ2/GLz13: e.g., \citealt{naidu2022, castellano2022}, and HD1: \citealt{harikane2022b}). 
In Section 4, we discuss the physical properties of $z\sim$11--17 candidates based on the full SED analysis results, and we also discuss the remaining low-$z$ possibility for each candidate in Section 5. 
A summary of this study is presented in Section 6. 
Throughout this paper, we assume a flat universe with 
$\Omega_{\rm m} =$ 0.3, 
$\Omega_\Lambda =$ 0.7, 
$\sigma_8 =$ 0.8, 
$H_0 =$ 70 km s$^{-1}$ Mpc$^{-1}$, 
and the Chabrier initial mass function (IMF) \citep{chabrier2003}. 
We place 2$\sigma$ upper limits for non detections unless otherwise specified. 
We take the cosmic microwave background (CMB) effect into account and correct the flux measurements at submm and mm bands, following the recipe presented by \cite{dacunha2013} (see also e.g., \citealt{pallottini2015, zhang2016, lagache2018}). 
\section{Observations \& Data} 
\label{sec:data}

\begin{figure*}
\includegraphics[trim=0cm 0cm 0cm 0cm, clip, angle=0,width=1\textwidth]{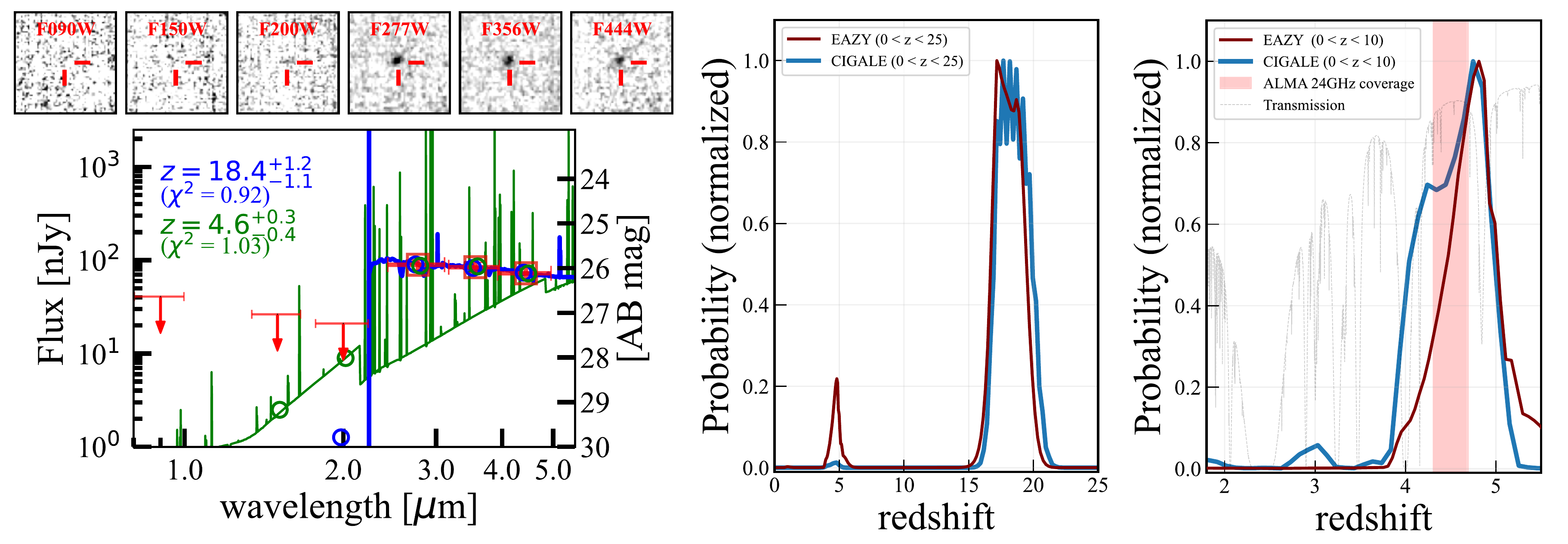}
\caption{
{\it Left:} 
The NIR SED of \targb. The red circles and arrows indicate the observed flux densities and 2$\sigma$ upper limits, respectively.
The blue and green curves and redshift labels represent the best-fit model SEDs and photometric redshifts by {\sc cigale} with the redshift range at $0<z<25$ and $0<z<10$, respectively. 
The blue and green open circles are predicted flux densities in the NIRCam filters based on the best-fit SEDs. 
The low-$z$ forced SED has a brighter submm flux by $>100$ times than the best-fit high-$z$ SED, expecting a $\sim10\sigma$ detection from the ALMA Band~7 observation (grey curve in Figure \ref{fig:nircam_alma}). 
The images on this panel present $2''\times2''$ NIRCam cutout images of \targb. 
{\it Middle:}
$P(z)$ from the SED fitting by {\sc eazy} (brown curve) and {\sc cigale} (light blue curve) with a redshift range at $0 < z< 25$. 
{\it Right:}
Same as the middle panel, but at $0 < z< 10$. 
The grey dashed line denotes the atmospheric transmission for \cii. 
The red shade indicates the \cii\ redshift range of $z=4.31$--4.69 covered by our ALMA Band~7 observations spanning 334--358~GHz with 3 frequency tunings, 
which is optimized to maximally cover the peak of the lower-redshift solution's $P(z)$ and avoid the significantly low atmospheric transmission. 
\label{fig:nircam}}
\end{figure*}

\subsection{JWST}
\label{sec:jwst_reduction} 
Stephan's Quintet, a group of five local galaxies was observed with NIRCam and MIRI in the JWST ERO program \citep{pontoppidan2022}. 
\targb\ falls in the coverage of NIRCam filters, but none of MIRI. 
The NIRCam images were taken in six bands: F090W, F150W, F200W, F277W, F356W, and F444W, covering 42~arcmin$^{2}$. The exposure time in each filter is $\sim$1200 sec. 
We use reduced and calibrated NIRCam imaging products that are publicly available\footnote{
\url{https://s3.amazonaws.com/grizli-v2/JwstMosaics/v4/index.html}}, 
and here we briefly explain the reduction and calibration procedure. 
The \jwst\ pipeline calibrated level-2 NIRCam imaging products were retrieved and processed with \texttt{grizli} pipeline \citep{brammer2021, brammer2022} in the same manner as \cite{bradley2022}.
The NIRCam photometric zero-point correction was applied with CRDS context \texttt{jwst\_0942.pmap}, including detector variations
\footnote{\url{https://github.com/gbrammer/grizli/pull/107}}. 
The derived photometric zeropoints are consistent with those derived by other teams with a \jwst\ ERS program \citep{boyer2022, nardiello2022}. 
While the consistent calibration results from a more recent calibration file of \texttt{jwst\_0989.pmap} have been confirmed within 3\% \citep{bradley2022}, we add a potential systematic uncertainty to the flux measurement by 10\% of the total flux in the following analyses to obtain secure results.  
The fully-calibrated images in each filter were aligned with the GAIA DR3 catalog \citep{gaia2021}, co-added, and drizzled at a 20~mas and 40~mas pixel scale for the short-wavelength (SW: F090W, F150W, F200W) and long-wavelength (LW: F277W, F356W, F444W) NIRCam bands, respectively. 

\subsection{ALMA}
\label{sec:alma_reduction}

\setlength{\tabcolsep}{5pt}
\begin{table*}
\caption{ALMA DDT Observation \& Data Properties for \targb}
\vspace{-0.6cm}
\label{tab:obs_summary}
\begin{center}
\begin{tabular}{cccccccccc}
\hline 
\hline
Freq. setup & Baseline & $N_{\rm ant}$ & Frequency & $T_{\rm int}$ & PWV  & beam & $<\sigma_{\rm line}>^{\dagger}$    & $\sigma_{\rm cont}^{\dagger}$  \\
 &    (m)       &     & (GHz)  & (min)  &  (mm)        &  ($''\times''$)    &  ($\mu$Jy/beam)    &  ($\mu$Jy/beam)          \\ \hline
Tuning1  & 15.1--629.3  & 43  & 334.02--337.90, 346.02--349.96  & 5.65   & 0.4 & $0.77\times0.46$ & 741 & 78.8   \\
Tuning2  & 15.1--629.3 & 42   & 338.02--341.90, 350.02--352.96  & 5.65   & 0.5 & $0.77\times0.46$ & 810 & 86.1   \\
Tuning3  & 15.1--629.3  & 42  & 342.02--345.90, 354.02--357.96  & 5.65   & 0.4 & $0.77\times0.46$ & 759 & 80.7  \\ \hline
Combined & \nodata & \nodata & $\sim$334--358 & 16.95 & \nodata & $0.77\times0.46$ & 770 & 45.0 \\
\hline \hline
\end{tabular}
\end{center}
\vspace{-0.2cm}
$\dagger$ Standard deviation of the pixels. For the cube, we show the average value from all channels in the 60-km~s$^{-1}$ data cube. 
\end{table*}

ALMA Band~7 observations were carried out on \targb\ on September 16th 2022 as a Cycle~8 DDT program (\#2021.A.00031.S, PI: S. Fujimoto).
The requested continuum sensitivity was achieved via three frequency setups ranging nearly 24-GHz wide over $\sim$334--358 GHz to maximize a chance of the \cii\ line detection at $z=4.31$--4.69 (red shade in the right panel of Figure~\ref{fig:nircam}) which covers around the peak of the redshift probability distribution $P(z)$ corresponding to the lower-redshift solution for \targb\ due to a lower-$z$ red galaxy with strong emission lines (Section~\ref{sec:photoz}).  
Each tuning was observed for 16 mins, resulting in a total of 48 mins including calibrations and overheads. 

The ALMA data were reduced and calibrated with the Common Astronomy Software Applications package versions 6.4.1.12 (CASA; \citealt{casa2022}) with the pipeline script in the standard manner. 
We imaged the calibrated visibilities with natural weighting, and a pixel scale of $0\farcs05$. 
For continuum maps, the {\sc tclean} routines were executed down to the 2$\sigma$ level with a maximum iteration number of 100,000 in the automask mode. 
For cubes, we adopted two common spectral channel bins of 15~km~s$^{-1}$ and 60~km~s$^{-1}$ and applied the {\sc tclean} routines with the same thresholds as the continuum map. 
The natural and tapered maps achieved full-width-half-maximum (FWHM) size of the synthesized beam of $0\farcs77\times0\farcs46$ with $1\sigma$ sensitivities for the continuum and the line in a 60-km s$^{-1}$ width channel of 45.0 $\mu$Jy and 770 $\mu$Jy beam$^{-1}$, respectively. 
We summarize the data properties of the continuum map and the cube in Table~\ref{tab:obs_summary}. 

\section{Analysis \& Results}
\label{sec:analysis}

\begin{figure*}
\includegraphics[trim=0cm 0cm 0cm 0cm, clip, angle=0,width=1\textwidth]{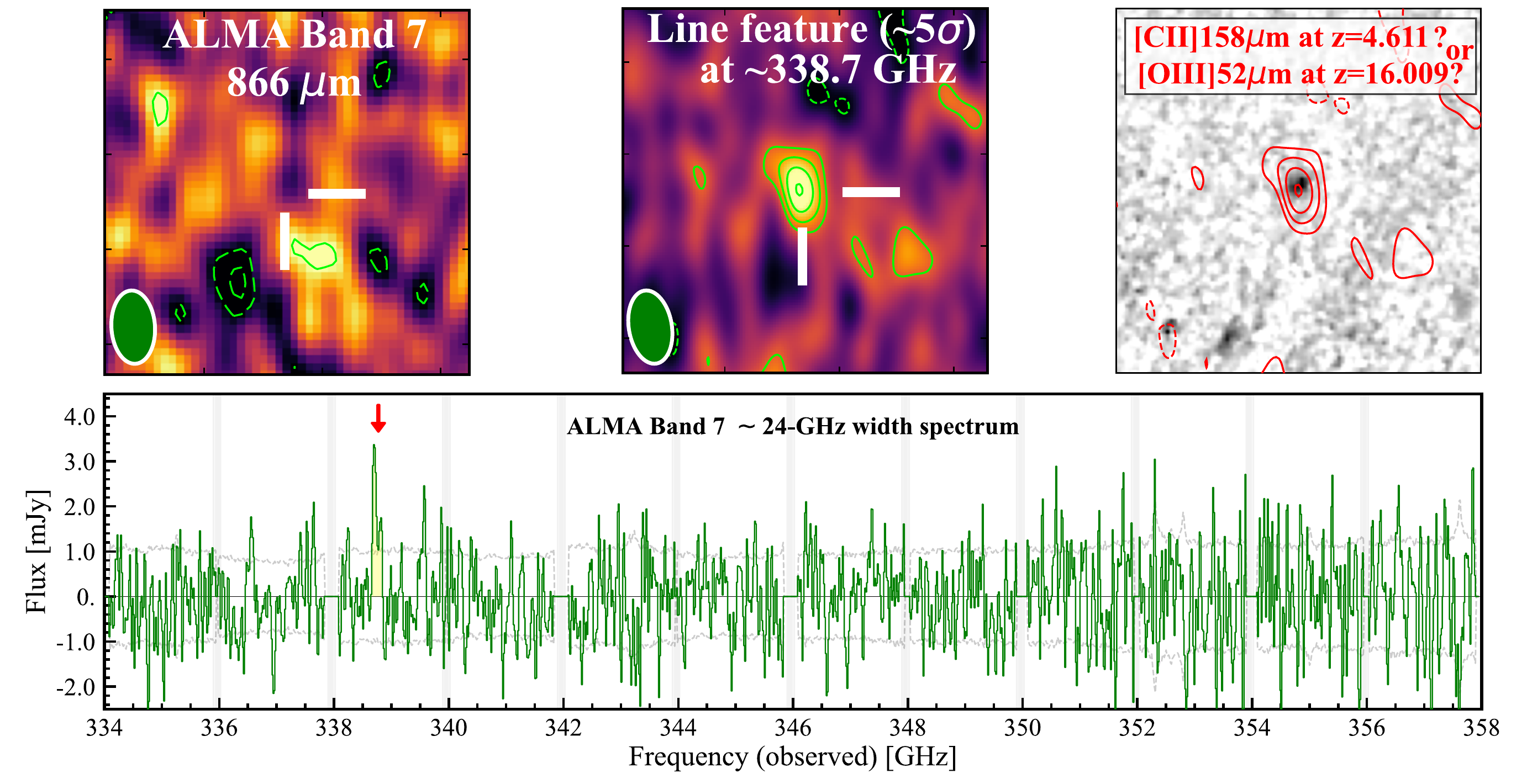}
\caption{
ALMA Band~7 observation results. 
{\it Top:}
ALMA $4''\times4''\times$ cutout of the continuum at 866~$\mu$m (left), the velocity-integrated map for the 5.1$\sigma$ line feature at around 338.7~GHz (middle), JWST/NIRCam F356W image (right). The dashed contours indicate the $-2\sigma$ and $-3\sigma$ levels, while the solid contours denote the $2\sigma$,  $3\sigma$,  $4\sigma$, and $5\sigma$ levels. 
The red contours overlaid on the F356W image indicate the line intensity in the moment-0 map. 
The green ellipse shows the ALMA synthesized beam. 
{\it Bottom:}
ALMA $\sim24$-GHz width spectrum (green) obtained from 3 frequency setups.
The grey dashed line denotes the 1$\sigma$ noise per channel. 
The grey shades show 120-MHz gaps between basedbands.
\label{fig:alma}}
\end{figure*}

\subsection{NIRCam photometry and redshift solutions}
\label{sec:photoz}

We use the \texttt{grizli} photometry catalog that is also publicaly available\footnote{
\url{https://s3.amazonaws.com/grizli-v2/JwstMosaics/v4/index.html}}. Briefly, the source fluxes in the NIRCam filters are evaluated with a circular aperture in $0\farcs36$ diameter and corrected to MAG\_AUTO. We correct the galactic dust reddening in the target direction.
In the left panel of Figure \ref{fig:nircam}, we present NIRCam cutouts and the \texttt{grizli} photometry for \targb. 
We confirm that \targb\ shows a clear dropout feature between F200W and F277W filters reported in \cite{harikane2023}, suggesting a Lyman-$\alpha$ break at $z\sim17$. 
We summarize the total flux measurements of \targb\ in Appendix.

We evaluate photometric redshifts ($z_{\rm phot}$) using {\sc cigale} \citep{burgarella2005, noll2009, boquien2019}. 
The fitting was performed in an identical fashion as in \cite{zavala2022}. 
In summary, we assume a delayed star-formation history (SFH): SFR($t$) $\propto t/\tau^{2}$ exp$(-t/\tau)$ with stellar models from \cite{bruzual2003}. Dust attenuation is also added following the dust attenuation law from \cite{calzetti2000} for the stellar continuum. The nebular emission (continuum + lines) is attenuated with a screen model and an SMC extinction curve \citep{pei1992}. During the SED fitting, the same $E(B-V)$ is used between stellar and nebular emission. Finally, the dust emission is re-emitted in the infrared modeled with \cite{draine2014} models. We list parameter ranges used in the fitting in the Appendix. 

Figure \ref{fig:nircam} summarizes the best-fit SED (left panel) and the probability distribution function $P(z)$ (middle panel) from {\sc cigale}. We obtain a photometric redshift of $z_{\rm phot}=18.4^{+1.2}_{-1.1}$,
supporting that \targb\ is a promising extremely high-redshift galaxy candidate \citep{harikane2023}. 
Note that \cite{harikane2023} report $z_{\rm phot}=16.7^{+1.9}_{-0.3}$ which is slightly lower than our estimate. This is because of the faint detection ($\sim2.4\sigma$) in the F200W filter in \cite{harikane2023}, while our F200W photometry is below the $1\sigma$ level, probably due to the difference in the reduction and calibration of the NIRCam data and the choice of the aperture size. 
We confirm the general consistency of the blue continuum color in the LW filters and the photometry in all NIRCam filters between ours and the latest one of \cite{harikane2023} (\textit{private comm.}) within the uncertainties. 

In $P(z)$, we also identify a non-zero probability at $z\sim5$. To assess the reasonable model for this secondary peak, we rerun {\sc cigale} with a limited redshift range of $0 < z < 10$ and show this $P(z)$ in the right panel of Figure \ref{fig:nircam}. 
We find that this best-fit low-redshift SED is composed of a red stellar continuum with strong rest-optical emission lines at $z_{\rm phot}=4.6^{+0.3}_{-0.4}$.  This model also well reproduces the NIRCam photometry, including the dropout feature in the F200W band. 
As shown in the middle panel of Figure \ref{fig:nircam},
although {\sc cigale} suggests 
a much lower likelihood at $z=4.6$
than the high-$z$ solution based on the Bayesian approach which applies the weights to all the models depending on the goodness-of-fit, the difference of the $\chi^{2}$ value from the high-$z$ solution ($\Delta \chi^{2}\equiv \chi^{2}_{\rm highz} -\chi^{2}_{\rm lowz}$) is only 0.11. 
This is because the optical emission lines of \oiii+H$\beta$ and H$\alpha$+\nii\ at $z\sim4$--5 fall exactly in the F277W and F356W filters, respectively, which boosts their broadband photometric fluxes to make them resemble the Lyman-$\alpha$ break feature for very specific cases among the model parameters. 
This is consistent with recent arguments discussed in the other the F200W dropout object known to be a similarly promising $z\sim17$ galaxy candidate, \targa\ \citep{zavala2022, naidu2022b}, and such a photometry boost effect in the NIR bands due to the strong emission lines have also been demonstrated by many authors before \jwst\ \citep[e.g.,][]{labbe2013, bowler2014, smit2014, smit2015, roberts-borsani2016}. 
In this forced lower-$z$ approach, we obtain a dusty galaxy solution with SFR = $50\,M_{\odot}$~yr$^{-1}$ and $M_{\rm star}=2.2\times10^{8}\,M_{\odot}$ with EW(\oiii+H$\beta$) = 450 ${\rm \AA}$, EW(H$\alpha$+\nii) = 240 ${\rm \AA}$, and a dust attenuation of the stellar continuum $E(B-V)=0.47$.  

We also carry out the SED fitting 
with {\sc eazy} (\citealt{brammer2008}), which performs the SED fitting to the observed photometry with a set of templates added in a non-negative linear combination. We use the default template set composed of the 12 \texttt{tweak\_fsps\_QSF\_12\_v3} templates derived from the Flexible Stellar Population Synthesis (FSPS) library \citep{conroy2009, conroy2010}. 
More details for {\sc eazy} are presented in \cite{kokorev2022}. 
Given our focus is to investigate the possibility that \targb\ may be a lower-$z$ red galaxy with strong emission lines suggested by {\sc cigale}, 
we modify an intermediate color star-forming template of \texttt{tweak\_fsps\_QSF\_12\_v3\_009} by boosting the emission line to EW(\oiii+H$\beta$) $\sim1100{\rm \AA}$ in a similar manner as \cite{labbe2022}. 
Note that this level of high EW(\oiii+H$\beta$) has been observed not only in young, early galaxies at $z\gtrsim6$ \citep[e.g.,][]{smit2014, endsley2021}, but also in lower-$z$ dusty objects including quasars \citep[e.g.,][]{finnerty2020, zakamska2003}. 
We set the redshift range to span from $0 < z < 25$, in steps of 0.01. 
We obtain the best-fit SEDs and $P(z)$ similar to those from 
{\sc cigale} in both cases: the redshift range at $0 < z < 25$ and $0 < z < 10$. 
Similar results are also obtained by using {\sc prospector} \citep{johnson2021} for \targa\ \citep{zavala2022}. 
The $P(z)$ from {\sc eazy} are also presented in the middle and right panels of Figure~\ref{fig:nircam}. 

We caution that the $\Delta \chi^{2}$ estimate is affected by the photometry measurements including the aperture choice and the aperture correction, the definition of the photometry uncertainties, the assumed parameter spaces of the model, and the implementations of each component (e.g., stellar population synthesis, nebular emission lines) in the model among the SED fitting codes. For instance, we conservatively add a potential systematic uncertainty in the NIRCam photometry by 10\% of the total flux to the original measurement uncertainty (Section \ref{sec:jwst_reduction}), where these additional errors can easily enhance the probability of lower-$z$ solutions \citep{naidu2022b}.  
Therefore, a different $\Delta \chi^{2}$ estimate from previous studies does not necessary weaken the robustness of the high-$z$ candidate selection in previous studies. 

\subsection{Dust continuum and FIR properties}
\label{sec:continuum}

\setlength{\tabcolsep}{2pt}
\begin{table}
\begin{center}
\caption{FIR properties of \targb}
\label{tab:fir_prop}
\vspace{-0.4cm}
\begin{tabular}{ccc}
\hline \hline
Redshift solutions                   & High-$z$ ($z\gtrsim16$) & Lower-$z$ ($z\sim5$) \\ \hline \hline
\multicolumn{3}{c}{Continuum} \\ \hline
$F_{\rm 866\mu m}$ [$\mu$Jy]         &  \multicolumn{2}{c}{$< 90.0$ (2$\sigma$)}  \\
$L_{\rm IR}$ [$L_{\odot}$] & $<1.2\times10^{12}$ & $<2.8\times10^{11}$         \\ 
SFR$_{\rm IR}$ [$M_{\odot}$~yr$^{-1}$] & $< 120$ & $< 28$ \\ \hline
\multicolumn{3}{c}{Line candidate} \\ \hline
Classification & \oiii\ 52~$\mu$m & \cii\ 158~$\mu$m  \\
S/N            & \multicolumn{2}{c}{5.1}    \\
Central frequency [GHz] & \multicolumn{2}{c}{$338.726 \pm 0.007$} \\
FWHM [km~s$^{-1}$] & \multicolumn{2}{c}{$118 \pm 20$} \\ 
$I_{\rm line}$ [Jy~km~s$^{-1}$]  & \multicolumn{2}{c}{$0.35\pm0.07$} \\
$z_{\rm line}$ &  $16.0089\pm0.0004$ &  $4.6108\pm0.0001$\\
$L_{\rm line}$ [$L_{\odot}$] & $(3.8\pm0.7)\times10^{9}$ & $(2.2\pm0.4)\times10^{8}$ \\
SFR$_{\rm line}^{\dagger}$ [$M_{\odot}$~yr$^{-1}$] & $\approx$ 30--130 & $\approx$ 20 \\ 
$M_{\rm dyn}^{\dagger\dagger}$ [$M_{\odot}$] & $\approx$ 1$\times10^{9}$ & $\approx$ 2$\times10^{9}$ \\ 

\hline \hline
\end{tabular}
\end{center}
\vspace{-0.2cm}
$\dagger$ Based on SFR--$L_{\rm line}$ relations in \cite{delooze2014} calibrated with local star-forming and metal-poor dwarf galaxies for the high-$z$ and lower-$z$ cases, respectively, 
where SFR$_{\rm [OIII]52}$ depends on the  \oiii52$\mu$m/\oiii88$\mu$m line ratio regulated by $n_{\rm e}$. Following the ratio of $\sim$1--5 ($n_{\rm e}\sim$100--3000~cm$^{-3}$) observed in local compact H{\sc ii} regions \citep{peeters2002}, we show the estimate with a range for SFR$_{\rm [OIII]52}$, where the lower side is comparable to the SFR estimate from the optical--mm SED fitting for the high-$z$ solution (Table \ref{tab:cigale_output}). 
\\
$\dagger\dagger$ Assuming the inclination angle of 45 deg and the diameter of $4\times r_{\rm e}$ measured with NIRCam \citep{ono2022}. 
\end{table}

The left panel of Figure \ref{fig:alma} shows the ALMA Band~7 continuum $4''\times4''$ image at 866 $\mu$m. 
The relevant pixels show negative counts. 
Based on the compact source size of \targb\ evaluated with NIRCam (effective radius $r_{\rm e}=0\farcs05$; \citealt{ono2022}), 
we assume that the emission is unresolved with the beam ($\sim0\farcs7$) in our ALMA map and place a 2$\sigma$ upper limit of 90.0~$\mu$Jy for the continuum emission based on the standard deviation of the map. Although we identify a weak signal ($\sim2\sigma$) with an offset by $\sim0{\farcs}8$, the offset is beyond the beam size, and we conclude that this nearby weak signal is a noise fluctuation irrelevant to \targb. 

In Figure \ref{fig:nircam_alma}, the red arrow represents the 2$\sigma$ upper limit from ALMA, and the grey curve indicates  the best-fit SED at $z=4.6$ based on NIRCam photometry with {\sc cigale} forced at $z<8$. The upper limit falls below the best-fit SED at $z=4.6$, strengthening the high-$z$ solution relative to a lower-redshift dusty galaxy with strong emission lines. We quantitatively investigate both scenarios based on the full SED analysis with the \jwst\ and ALMA photometry in Section \ref{sec:sed}.

We evaluate the upper limit of the IR luminosity ($L_{\rm IR}$) and obscured SFR (SFR$_{\rm IR}$) for \targb\ based on the following dust temperature ($T_{\rm d}$) estimates. 
First, we extrapolate the best-fit redshift evolution model of $T_{\rm d}$ following the decrease of the gas depletion time scale ($t_{\rm depl}$) derived in \cite{sommovigo2022}, and obtain $T_{\rm d}=90$~K at $z=18.0$.\footnote{We assume the gas-phase metallicity of $Z=0.1\,Z_{\odot}$ and the effective dust attenuation optical depth of $\tau_{\rm eff}=-$ln$T$ with $T=0.9$.} 
Although the extrapolation out to $z\sim18$ is challenging, 
we note that $t_{\rm depl}$ is likely very short in \targb\ due to a very compact source size of $r_{\rm e}=140^{+90}_{-60}$~pc and a very high surface SFR density of $\Sigma_{\rm SFR}\sim180\,M_{\odot}$~yr$^{-1}$~kpc$^{-2}$ from the rest-frame UV measurements with NIRCam based on the high-$z$ solution \citep{ono2022}.  

Second, we calculate the radiative equilibrium model with a clumpy ISM distribution in the same manner as \citep{inoue2020, fudamoto2022a}. Assuming the same rest-FIR continuum size as the F277W measurement, we obtain a lower limit of $T_{\rm d}\sim80$~K
\footnote{
We obtain the lower limits of 96~K and 81~K with the 2$\sigma$ and 3$\sigma$ upper limit of the dust continuum at 866~$\mu$m, respectively, where we adopt the lower limit of 80~K, given uncertainties from the assumptions in the model calculation. 
}. 
Based on the agreement from these two approaches,  
we adopt a single modified black body (MBB) with $T_{\rm d}=90$~K and the dust spectral index $\beta_{\rm d}=2.0$\footnote{
This is the same assumption as \cite{sommovigo2022} and \cite{fudamoto2021}. 
} and infer $L_{\rm IR} < 1.2\times10^{12}\,L_{\odot}$ and SFR$_{\rm IR} <120\,M_{\odot}$~yr$^{-1}$,\footnote{We assume SFR [$M_{\odot}$~yr$^{-1}$] =1.0$\times10^{-10}\,L_{\rm IR}$ [$L_{\odot}$]}
if \targb\ is truly an ultra high-redshift object at $z\sim18$.
We caution that the CMB temperature at $z=18$ reaches $\sim50$~K. Thus, a lower $T_{\rm d}$ assumption of e.g., 60~K also provides a similar upper limit after the CMB correction. 
In the case that \targb\ is a lower-$z$ object at $z\sim4.6$ (Section~\ref{sec:photoz}), we obtain $T_{\rm d}=49$~K from the same $T_{\rm d}(z)$ model from \cite{sommovigo2022}, which satisfies again the lower limit of $T_{\rm d}>30$~K estimated from the radiative equilibrium model \citep{inoue2020, fudamoto2022a}. 
From the same single MBB with $T_{\rm d}=49$~K, we infer $L_{\rm IR} < 2.8\times10^{11}\,L_{\odot}$ and SFR$_{\rm IR} < 28\,M_{\odot}$~yr$^{-1}$. 
This rules out the possibility that \targb\ is a lower-$z$ DSFG with SFR $=$ 50 $M_{\odot}$~yr$^{-1}$, which is suggested by the forced low-$z$ SED before ALMA (Section \ref{sec:photoz}).  
We further investigate the full SED properties including the new ALMA photometry in Section \ref{sec:sed}. 
We summarize our estimates of the FIR properties in Table~\ref{tab:fir_prop}.

\subsection{ALMA $\sim$24-GHz width line scan in Band~7}
\label{sec:scan}

To gain further insight into the redshift of this source, we analyze the $\sim$24-GHz wide spectrum in Band~7 to search for a serendipitous line detection. 
The frequency setup is optimized to cover the peak of $P(z)$ at $z=4.31$--4.69 with \cii\ 158$\mu$m emission line and avoid the significantly low atmospheric transmission, which is summarized in the right panel of Figure~\ref{fig:nircam}.
Note that there is a $\sim$120~MHz gap between each baseband. However, this frequency gap corresponds to $\sim100$~km~s$^{-1}$, which is narrower than typical \cii\ line widths of $\sim300$--1200 km~s$^{-1}$ among high-$z$ DSFGs \citep[e.g.,][]{carilli2013} and thus does not much affect our \cii\ line identification from typical DSFGs. 

In the bottom panel of Figure~\ref{fig:alma}, we show the Band~7 spectrum of \targb\ from the 15-km~s$^{-1}$ channel cube. 
Given the compact source size, we assume the emission unresolved and extract the spectrum with the mean pixel count within a $0\farcs2$ diameter in the Jy~beam$^{-1}$ unit.
In the spectrum, we identify a line feature at around $338.7$ GHz, where the positive signals continue in 12 consecutive channels. 
We produce a velocity-integrated (moment-0) map and obtain a significance level of 5.1$\sigma$ at the peak pixel in the moment-0 map. From a single Gaussian fit to the spectrum, we evaluate the line width full-width-half-maximum (FWHM) to be $118\pm20$~km~s$^{-1}$, a line intensity of $I_{\rm line}=0.35\pm0.07$ Jy~km~s$^{-1}$, and a central frequency at $338.726\pm0.007$~GHz. 

In Figure~\ref{fig:alma}, we show the moment-0 map (top middle) and the contour of the line intensity overlaid on the NIRCam/F356W map (right). 
The peak position of the line intensity exactly matches with the NIRCam source position, suggesting that this is one of the most promising line features among the recent ALMA observations for $z>11$ candidates, where multiple tentative ($\sim4\sigma$) features have been identified with small spatial offsets \citep{harikane2022b, bakx2022, yoon2022}. 
We find that other weak positive signals appeared in the spectrum (e.g., 336.6~GHz \& 339.5~GHz) always show the peak and morphology in the moment-0 map not well aligned with the NIRCam source position with spatial offsets ($\gtrsim0\farcs2$), being the most likely noise, in contrast to the 338.7-GHz line feature. 
To understand the noise properties more, 
we also generate a data cube with a 162-km~s$^{-1}$ channel width, which consists of a total of 5,701,600 voxels based on the number of channels and the pixels of the cube. We estimate the number of similarly bright ($>2$~mJy) noise voxels in this data cube and find that the chance probability is estimated to be $\sim2\%$ to identify a noise peak with $> 2$~mJy within one beam-radius search volume. 

To further address the reliability of this line candidate, we also run a blind line search algorithm of {\sc Findclump} implemented in a Python library of {\sc Interferopy} \citep{interferopy} for observational radio--mm interferometry data analysis\footnote{
\url{https://interferopy.readthedocs.io/en/latest/index.html}
}. 
For this analysis, we also produce data cubes with different channel widths of 20-km~s$^{-1}$ and 30-km~s$^{-1}$ and find that the line candidate is always recovered with S/N = 4.7--5.3 in the blind search algorithm regardless of the choice of the data cube with different channel widths. 
From the histograms of the positive and negative detections, the fidelity\footnote{
Fidelity (S/N) $\equiv$ [$N$(positive) - $N$(negative)]/ $N$(positive), where $N$ is the number of detection with a given S/N.
} at the line S/N is estimated to be $\sim50$\%. Note that this is a blind search approach in the entire data cube. Therefore, the realistic fidelity at the source position is much higher than 50\%. 
We conclude that the fidelity of this line candidate is  (conservatively) at least 50\%, and the most likely $\sim$98\% from the above estimate based on the prior information of the target position.
Given that no significant emission is detected in both continuum and each channel in the cube, we also produce the dirty cubes (i.e. applying no CLEAN) and confirm the same results. 
In the Appendix, we show the fidelity curve estimated from the positive and negative histograms as a function of S/N. 
Table \ref{tab:fir_prop} summarizes the properties of the line candidate. 

\subsection{Line interpretation}
\label{sec:line}

Based on the two redshift solutions of 
$z_{\rm phot}=18.4^{+1.2}_{-1.1}$ and $z_{\rm phot}={4.6}^{+0.3}_{-0.4}$ (Section \ref{sec:photoz}), the possible interpretation for the line is 
\oiii~52~$\mu$m at $z=16.0089\pm0.0004$ or 
\cii~158~$\mu$m at $z=4.6108\pm0.0001$. 
Although the middle panel of Figure~\ref{fig:nircam} suggests $P(z>16)$ is much higher than that of the lower-$z$ solution, the F200W filter starts including the flux from the red side of the Lyman-alpha break at $z\lesssim17$, which makes $P(z)$ at $z=16.0$ not as high as the redshift solutions at $z\sim17$--19. 
From $P(z)$, the likelihoods at $z=16.0$ and $z=4.6$ are almost comparable and thus difficult to conclude which is more likely only from this aspect.
Although the upper limit of the dust continuum rules out the possibility of the lower-$z$ DSFG with SFR $\gtrsim30\,M_{\odot}$~yr$^{-1}$ (Section \ref{sec:continuum}), we further discuss the remaining possibilities of the low-$z$ solution in Section \ref{sec:low-z}. 
We also explore the possibility of CO(3-2) at $z=0.0208\pm0.0002$ in Appendix, which we conclude unlikely. 
Therefore, we examine both interpretations in this subsection. 

In the $z=4.6$ case, we estimate a \cii\ line luminosity of $L_{\rm [CII]}=(2.2\pm0.4)\times10^{8}\,L_{\odot}$ and SFR of $\approx20\,M_{\odot}$~yr$^{-1}$ based on the SFR--$L_{\rm [CII]}$ relation calibrated among local star-forming galaxies \citep{delooze2014}.
This yields the $L_{\rm [CII]}/L_{\rm IR}$ ratio of $\gtrsim8\times10^{-4}$, which falls in the typical range of $\sim10^{-2}$--$10^{-4}$ observed among dusty star-forming galaxies at $z\sim$0--6 \citep[e.g.,][]{diaz-santos2013, gullberg2015}. 
In the $z=16.0$ case, we calculate an \oiii\ 52$\mu$m line luminosity of $L_{\rm [OIII]52}=(3.8\pm0.7)\times10^{9}\,L_{\odot}$. 
Based on the SFR--$L_{\rm [OIII]88}$ relation estimated among local metal-poor galaxies \citep{delooze2014} and the typical line ratio of \oiii52$\mu$m and \oiii88$\mu$m lines of $\sim1$--5 observed in local compact H{\sc ii} regions \citep{peeters2002}, we evaluate the SFR value to be $\approx30$--130 $M_{\odot}$~yr$^{-1}$. 
Although systematic uncertainties remain in the application of these empirical relations, we confirm that our line-based SFR estimates are consistent with the upper limits of SFR$_{\rm IR}$ from the dust continuum in both cases. 
We caution that the high \oiii52$\mu$m/88$\mu$m ratio\footnote{The \oiii52$\mu$m/88$\mu$m line ratio is regulated by electron density due to difference of their critical densities, and not much affected by metallicity and ionization parameter \citep[e.g.,][]{jones2020,yang2021}} of $\sim5$ indicates a high electron density of $n_{\rm e}\sim 3000$ cm$^{-3}$ which exceeds the critical density of \oiii\ 88$\mu$m. It is thus unclear whether the assumed SFR--$L_{\rm [OIII]88}$ relation, which is also affected by the metallicity and ionization parameter, is validated in this high $n_{\rm e}$ regime. A dedicated analysis will be necessary in a separate paper.

Following the method outlined in \cite{wang2013}\footnote{In approximation, the dynamical mass is given by $M_{\rm dyn}=1.16\times10^{5}V_{\rm circ}D$, where $D$ is the diameter and $V_{\rm circ}$ is circular velocity. $V_{\rm circ}$ is also given by $V_{\rm circ}=1.763\sigma_{\rm line}/\sin(i)$, where $i$ is inclination angle and $\sigma_{\rm line}$ is the velocity dispersion of the line.
We assume the inclination of 45 deg and $D=4\times r_{\rm e}$ from the NIRCam observation.}, 
we also estimate the dynamical mass of $M_{\rm dyn}\approx2\times10^{9}M_{\odot}$ and $\approx1\times10^{9}M_{\odot}$ in the $z=4.6$ and $z=16.0$ cases, respectively.
In the $z=4.6$ case, the $M_{\rm star}$ value is estimated to be $2\times10^{8}\,M_{\odot}$ in our forced low-$z$ SED analysis (Section \ref{sec:photoz}), and thus the $M_{\rm star}/M_{\rm dyn}$ ratio is about 10\%. Also with the upper limit of SFR$_{\rm IR}$, this suggests that \targb\ is a moderately star-forming, very gas-rich system (gas fraction $\sim$90\%) at $z=4.6$, which is consistent with the recent ALMA results for main-sequence galaxies surrounded by rich metal-rich gas reservoir at $z\sim4$--7 \citep[e.g.,][]{mirka2020, fujimoto2019, fujimoto2020b, fujimoto2021}. 
In the high-$z$ scenario, \cite{harikane2023} evaluate $M_{\rm star}=7.0^{+50.8}_{-4.8}\times10^{8}\, M_{\odot}$ for \targb, which satisfies $M_{\rm star} \leq M_{\rm dyn}$. 
Assuming that $M_{\rm dyn}$ is dominated by the molecular gas and stellar masses, the above estimates indicate a low gas fraction ($\approx0.3$) in the high-$z$ scenario, which is likely consistent with the decreasing trend of the gas fraction with increasing stellar mass \citep[e.g.,][]{tacconi2013}. 
Note that the structure formation model with Planck cosmology ({\sc universemachine}; \citealt{behroozi2020}), the most massive dark matter halos at $z=16.0$ is calculated to be $M_{\rm halo}\sim8\times10^{9}\,M_{\odot}$. 
Thus, the $M_{\rm star}/M_{\rm halo}$ ratio can be still $\sim0.09$ that satisfies the upper boundary from the cosmic baryon fraction of 0.16. 
One note is that such a high stellar-to-halo-mass ratio implies a significantly high star-formation efficiency. We further discuss the validity of the high-$z$ solution in Section \ref{sec:highz}.

Based on these results, both interpretations are possible, and it is challenging to conclude which is more likely with the current data sets.
Once the line feature is confirmed, 
the low-$z$ solution at $z=4.61$ will be verified with the \jwst/NIRSpec follow-up by targeting the strong rest-frame optical emission lines that cause the dropout feature between the F200W and F277W filters. 
In fact, this is the case of another extremely high-$z$ galaxy candidate, initially estimated at $z\sim17$ \citep[e.g.,][]{donnan2022}, which has been subsequently spectroscopically confirmed to be at $z=4.91$ \citep{arrabal-halo2023a}.
If we do not detect any emission lines from NIRSpec, ALMA follow-up observations for the \oiii\,88~$\mu$m line will be a plausible approach to spectroscopically confirm the high-$z$ solution at $z=16.01$, since the bright rest-frame optical emission lines (e.g., \oiii5007, H$\beta$) shift out of the spectral window of NIRSpec at $z\gtrsim11$. 
We summarize the properties of the line candidate in both cases in Table \ref{tab:fir_prop}. 

\subsection{\jwst+ALMA joint SED analysis}
\label{sec:sed}

\setlength{\tabcolsep}{10pt}
\begin{table*}
\begin{center}
\caption{Summary of UV luminous $z\sim11$--17 galaxy candidates  observed with ALMA}
\label{tab:recent_alma}
\vspace{-0.4cm}
\begin{tabular}{cccccccc}
\hline 
Source Name & R.A. & Dec. & $z_{\rm phot}^{\rm literature}$ & F444W  & $\lambda_{\rm obs}^{\rm ALMA}$ & $F_{\rm ALMA}$ & Ref. \\
          &  (deg) & (deg) & & (mag) &  (mm)  & ($\mu$Jy) & \\ 
(1) & (2) & (3) & (4) & (5) & (6) & (7) & (8) \\ \hline 
S5-z17-1& 339.015969& 33.904624& $16.7^{+1.9}_{-0.3}$ &$26.75\pm0.17$ &0.87 &$<$ 90.0 & This work  \\
GHZ1/GLz11 &3.498988 & $-30.324759$ & $10.4^{+0.2}_{-0.7}$ &$26.31\pm0.11$ &1.21 & ($15.6\pm5.8$) & Y22 \\ 
GHZ2/GLz13  &3.511923 & $-30.371859$ & $12.4^{+0.1}_{-0.3}$ &$26.67\pm0.11$ & 1.02 & $<$ 7.2 & B22, P22 \\ 
\multirow{2}{*}{HD1} & \multirow{2}{*}{150.463792} & \multirow{2}{*}{2.547222} & \multirow{2}{*}{$15.2^{+1.2}_{-2.7}$} & \multirow{2}{*}{$24.67\pm0.30$} & 1.27 & $< 16.0$ & H22  \\ 
 \multicolumn{5}{c}{} & 2.17 & $< 10.4$ & K22  \\  \hline
\end{tabular}
\end{center}
\vspace{-0.4cm}
\tablecomments{
(1) Source name in literature. 
(2) Right ascension. (3) Declination. 
(4) Photometric redshift estimate in literature (S5-z17-1: \citealt{harikane2023}, GHZ1/GLz11 \& GHZ2/GLz13: \citealt{naidu2022}, HD1: \citealt{harikane2022b})
(5) Our total magnitude estimate in NIRCam/F444W filter with 1$\sigma$ errors. The potential systematic uncertainty is added by 10\%.
For HD1, we show the total magnitude estimate of {\it Spitzer} IRAC ch2 in \cite{harikane2022b}. 
(6) Observed wavelength in the ALMA observation based on the center sky frequency.   
(7) Submm--mm photometry from our and recent ALMA observations. The upper limit is placed at the 2$\sigma$ level, and a tentative 2.6$\sigma$ emission is reported in GHZ1/GLz11 \citep{yoon2022}.  
(8) Reference of the ALMA observation (Y22: \citealt{yoon2022}, B22: \citealt{bakx2022}, P22: \citealt{popping2022}, H22: \citealt{harikane2022b}, and K22: \citealt{kaasinen2022}). 
}
\end{table*}

\begin{figure*}
\includegraphics[trim=0cm 0cm 0cm 0cm, clip, angle=0,width=1.
\textwidth]{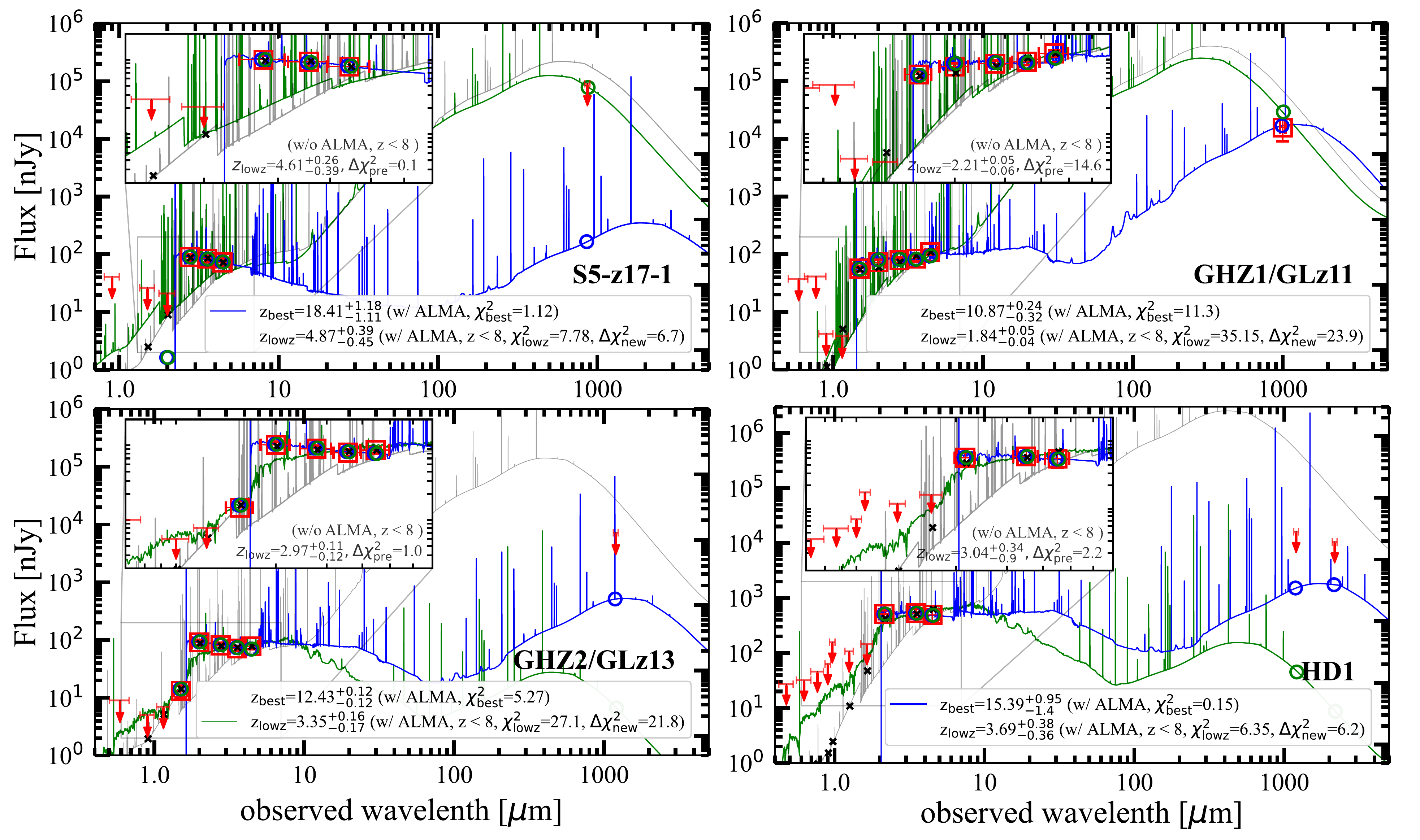}
\caption{
Optical to mm SED of the $z\sim11$--17 candidates so far observed with ALMA. 
The red open squares and arrows indicate the observed flux densities and the $2\sigma$ upper limits, respectively. 
The blue curve is the best-fit SED with the optical--mm photometry, showing that the high-$z$ solution is favored in all candidates. 
For comparison, the green and grey curves indicate the best-fit SEDs forced at $z < 8$ with and without the ALMA photometry. The lower-$z$ IR-bright objects, implied from the grey curves, are ruled out by the new constraints from ALMA in all candidates. 
The $\Delta\chi^{2}_{\rm pre}$ and $\Delta\chi^{2}_{\rm new}$ values in the labels indicate the the difference of the $\chi^{2}$ values between the forced low-$z$ and the best-fit high-$z$ solutions without and with ALMA data, respectively.
All candidates show $\Delta\chi^{2}_{\rm new}$ increased from $\Delta\chi^{2}_{\rm pre}$, indicating that their high-$z$ solutions are much strengthened with the new ALMA photometry. 
The open circles and black crosses denote the predicted photometry from the SEDs in each filter and ALMA band. 
\label{fig:nircam_alma}}
\end{figure*}

The non-detection of the dust continuum from \targb\ is reminiscent of recent ALMA results in other three UV-bright galaxy candidates at $z\sim11$--13: GHZ1/GLz11 , GHZ2/GLz13, and HD1 \citep{harikane2022b, bakx2022, popping2022, yoon2022, kaasinen2022, yoon2022}. 
GHZ1/GLz11 and GHZ2/GLz13 were also identified in the early \jwst\ data from the GLASS field \citep{treu2022} from different teams \citep[e.g.,][]{castellano2022, naidu2022, donnan2022, harikane2023}. No robust dust continuum is detected in follow-up deep 1-mm observations with a total of $>10$-hrs observing time for both candidates \citep{bakx2022, popping2022, yoon2022}, while a tentative ($2.6\sigma$) detection is reported in GHZ1/GLz11 \citep{yoon2022}.
HD1 was found as a remarkably bright ($M_{\rm UV}\sim-24$) galaxy candidate at $z\sim13$ in a systematic search over a 2.3~deg$^{2}$ area in ground-based telescopes and {\it Spitzer} data \citep{harikane2022b}. Similarly deep ALMA 1-mm and 2-mm band observations have been carried out, showing no dust continuum detection in both ALMA observations \citep{harikane2022b, kaasinen2022}.
These results may imply a low possibility of contamination from lower-$z$ dusty star-forming galaxies with strong emission lines among the high-$z$ candidates at $z\sim$11--17 recently identified and observed with ALMA. In Table \ref{tab:recent_alma}, we summarize \targb\ and these three UV-bright high-$z$ galaxy candidates so far observed with ALMA.

To further investigate the high-$z$ ($z\gtrsim11$) and the lower-$z$ scenarios for all these candidates, we perform SED fitting to the optical--mm photometry using {\sc cigale} \citep{burgarella2005, noll2009, boquien2019}. We adopt the same assumptions in the fitting described in Section~\ref{sec:photoz}. We use the public \textit{grizli} catalog for GHZ1/GLz11 and GHZ2/GLz13, where the \jwst\ data reduction, calibration, and photometry are processed in the same manner as \targb\ (Section \ref{sec:jwst_reduction}). We also use the photometry of the \hst/ACS--WFC3 images in the catalog, including the latest ACS data taken as part of a DDT program (\#17231, PI: T. Treu), which is processed using the \texttt{grizli} pipeline in the same manner as \cite{kokorev2022}. 
We list the \jwst\ and \hst\ photometry of GHZ1/GLz11 and GHZ2/GLz13 in the Appendix. 
The optical--NIR photometry of HD1 is taken from \cite{harikane2022b}. The ALMA photometry measurements of GHZ1/GLz11, GHZ2/GLz13, and HD1 are taken from the previous studies \citep{bakx2022, popping2022, yoon2022, harikane2022b, kaasinen2022}. 
We use the photometry with the $1\sigma$ error also for the  measurements below the $2\sigma$ upper limits. When the literature only provides the upper limit, we set zero flux with the 1$\sigma$ error in those non-detection bands. To maintain the same detection thresholds among different wavelengths, we use the 2.6$\sigma$ detection in the ALMA 1-mm band in GHZ1/GLz11. 

\setlength{\tabcolsep}{5pt}
\begin{table*}[t!]
\begin{center}
\caption{Physical properties of $z=11$--17 candidates from optical--mm SED fitting}
\label{tab:cigale_output}
\vspace{-0.4cm}
\begin{tabular}{cccccccc}
\hline  \hline
Source Name & $z_{\rm best}$ ($\chi^{2}$) & $z_{\rm lowz}$ ($\chi^{2}$) & $\Delta\chi^{2}$ &$M_{\rm UV}$ & $\beta_{\rm UV}$ & SFR$_{\rm 10Myr}$ & $M_{\rm star}$  \\
  &   &   &  & (mag) &    & ($M_{\odot}$~yr$^{-1}$) & ($10^{9}\,M_{\odot}$)  \\ 
      & (1) & (2) & (3) & (4) & (5) & (6) & (7) \\ \hline 
S5-z17-1 & $18.41^{+1.18}_{-1.11}$ (1.12) &$4.45^{+0.46}_{-0.52}$ (7.84) &6.71 &$-21.87\pm0.11 $&$-2.04\pm0.05 $&$23^{+8}_{-4}$ &$1.1^{+0.7}_{-0.6}$  \\
GHZ1/GLz11 & $10.87^{+0.24}_{-0.32}$ (11.3$^{\dagger}$) & $1.84^{+0.05}_{-0.04}$ (35.15) & 23.85 &$-21.03\pm0.12 $&$-2.29\pm0.02 $&$15^{+3}_{-2}$ &$1.4^{+0.5}_{-0.4}$  \\
GHZ2/GLz13 & $12.43^{+0.12}_{-0.12}$ (5.27$^{\dagger}$) & $3.35^{+0.16}_{-0.17}$ (27.10) & 21.83 &$-21.35\pm0.07 $&$-2.45\pm0.01 $&$13^{+2}_{-1}$ &$0.8^{+0.3}_{-0.4}$  \\
HD1 & $15.39^{+0.95}_{-1.40}$ (0.15) & $3.69^{+0.38}_{-0.36}$ (6.35) & 6.20 &$-23.64\pm0.18 $&$-2.22\pm0.03 $&$101^{+24}_{-16}$ &$5.4^{+2.8}_{-2.7}$   \\ 
\hline \hline
\end{tabular}
\end{center}
\vspace{-0.4cm}
\tablecomments{
(1) Photometric redshift with the best-fit SED at $0 < z < 25$. 
The $\chi^{2}$ value is shown in parentheses.  
(2) Photometric redshift with the best-fit SED forced at $0 < z < 8$. 
The $\chi^{2}$ ($z_{\rm lowz}$) value is shown in parentheses.
(3) Difference of the $\chi^{2}$ values between the best-fit SEDs at $z_{\rm best}$ and $z_{\rm lowz}$. 
(4--7) Physical properties in the high-$z$ solutions based on $z_{\rm best}$: (4) Absolute UV magnitude, 
(5) UV continuum slope measured by a single power-low fit to the continuum component in the best-fit SED over rest-frame 1400--2500 {\rm \AA} in a similar manner as \cite{nanayakkara2022}, 
(6) Average SFR over 10 Myr, 
(7) Stellar mass.  
}
$^{\dagger}$ The best-fit SEDs with smaller $\chi^{2}$ values are obtained in the literature, while our measurements include the new ALMA photometry which affects the best-fit parameter space and the $\chi^{2}$ value. 
\end{table*}

In Figure~\ref{fig:nircam_alma}, we show the best-fit SED (blue curve) with the optical--mm photometry (red symbols). 
For comparison, we also show the best-fit SED forced at $0< z < 8$ with (green curve) and without the ALMA photometry (grey dashed curve). 
For every candidate, we find that the best-fit SED from the optical--mm photometry not only favors the high-$z$ solution at $z\gtrsim11$. 
 Moreover, the ALMA photometry always falls below the grey dashed curve, suggesting that the possibility of lower-$z$ IR-bright DSFGs are ruled out. 
Still, the possibility of lower-$z$, IR-faint red objects might remain, which corresponds to the best-fit SED forced at low-$z$ with the ALMA photometry (green curves). 
 In the inset labels, we also present the $\Delta \chi^{2}$ values of the forced low-$z$ solutions from the best-fit high-$z$ solutions in the SED analysis before ($\Delta\chi^{2}_{\rm pre}$) and after including the ALMA photometry ($\Delta\chi^{2}_{\rm new}$). 
 We find that the $\Delta \chi^{2}$ value increases in every candidate out to $\sim$6--27 (i.e. $\Delta\chi^{2}_{\rm pre} < \Delta\chi^{2}_{\rm new}$; the addition of the ALMA non-detection increases the likelihood of the high-redshift solution relative to the low-redshift solution), satisfying the criterion of $\Delta\chi^{2}>4.0$, corresponding to a $2\sigma$ level, used in previous studies \citep[e.g.,][]{bowler2020, harikane2022b, donnan2022, finkelstein2023}. 
These results suggest that the lower-$z$ IR-faint red objects are also unlikely supported, although $\Delta\chi^{2}$ values may change with different SED codes and assumptions (e.g., high $T_{\rm dust}$). 
We further discuss the remaining possibilities of the lower-$z$ solution in Section \ref{sec:low-z}.

\section{Blue Monsters at \lowercase{$z$} $\sim$ 11--17}
\label{sec:discussion}

\subsection{Presence of UV bright galaxies out to $z\sim17$}
\label{sec:highz}

\begin{figure*}
\includegraphics[trim=0cm 0cm 0cm 0cm, clip, angle=0,width=1\textwidth]{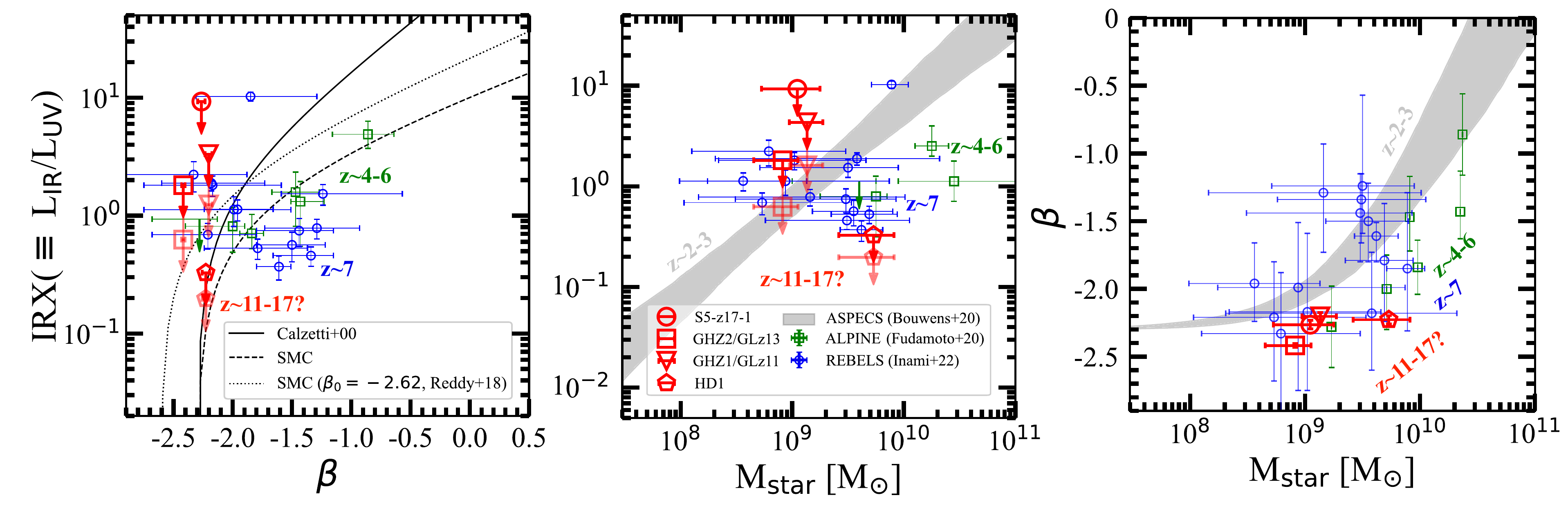}
\caption{
Comparison of IRX, $\beta_{\rm UV}$, and $M_{\rm star}$ properties with other high-$z$ star-forming galaxies constrained from recent large ALMA surveys of ASPECS at $z\sim2$--3 \citep{walter2016}, ALPINE at $z\sim4$--6 (green square; \citealt{lefevre2020}), and REBELS $z\sim7$ (blue square; \citealt{bouwens2021}). 
Note that ASPECS and ALPINE results are taken from the stacking results \citep{bouwens2020,fudamoto2020}, while REBELS results are taken from the individual results based on detection \citep[e.g.,][]{inami2022, sommovigo2022}. 
The red symbols represent the UV-bright ($M_{\rm UV}\approx[-24:-21]$) high-$z$ candidates at $z\sim$11--17 constrained from our optical--mm SED analysis. The solid and light red symbols are estimated from the $T_{\rm d}-z$ relation of \cite{sommovigo2022} and a constant assumption of $T_{\rm d}=$ 50~K, respectively. 
The upper limits are placed at the $2\sigma$ level, while we place the 2.6$\sigma$ upper limit for GHZ1/GLz11 that has been reported to have a tentative (2.6$\sigma$) continuum emission \citep{yoon2022}. 
{\it Left:} IRX--$\beta_{\rm UV}$ relation. The solid and dashed curves indicate the relations derived with the dust attenuation of SMC and \cite{calzetti2000}, respectively. 
The dotted curve shows the relation derived with the SMC dust attenuation and bluer intrinsic $\beta_{\rm UV}$. 
{\it Middle \& Right:} IRX--$M_{\rm star}$ and $M_{\rm star}$--$\beta_{\rm UV}$ relations from middle to right. 
The black shade shows the 1$\sigma$ range of the best-fit relations at $z\sim$2--3 estimated in \cite{bouwens2020}. 
Although both UV-bright $z\sim$11--17 candidates and the REBELS sources dominate the bright-end of UVLF and the similar $M_{\rm star}$ range at these redshifts, the former is generally bluer and lower IRX. 
\label{fig:irx}}
\end{figure*}

Owing to our and recent deep ALMA observations, the high-$z$ solutions at $z\sim11$--17 are all favored in the UV bright high-$z$ candidates of S5-z17-1, GHZ1/GLz11, GHZ2/GLz13, and HD1 (Section \ref{sec:sed}). 
In particular, the high-$z$ solution obtained from \targb\  suggests the presence of the remarkably UV-bright ($M_{\rm UV}=-21.9$) object at $z\sim17$, just $\sim200$~Myr after the Big Bang. 
This UV luminosity is comparable to that of GN-z11 \citep{oesch2016}, making S5-z17-1 the second most luminous object at $z>11$ after HD1 ($M_{\rm UV}=-23.6$). Such an identification in the small survey volume among the early \jwst\ observations could present a challenge to the current models of early galaxy formation and potentially even the underlying $\Lambda$CDM cosmological framework \citep[e.g.,][]{steinhardt2016, steinhardt2022, mason2022, boylan2022, lovell2022, menci2022}. 
As discussed in \cite{naidu2022b}, no theoretical UVLF or empirical extrapolation can be close to matching with its presence, except for a 100\% instantaneous star-formation efficiency coupling with the dark matter halo mass function, while the star-formation efficiency measured at $z\sim6$--10 is typically $<10\%$ \citep[e.g.,][]{finkelstein2015b, tacchella2018, stefanon2021}. 

\cite{harikane2023} discuss three possible scenarios (see also \citealt{inayoshi2022}) for the presence of a remarkably UV bright object even out to $z\sim17$: A) no star-formation suppression, B) presence of active galactic nuclei (AGN), and C) pop III like stellar population with a top-heavy IMF. 
For the scenario A), 
recent numerical studies of star cluster formation from compact giant molecular clouds also indicate high star-formation efficiency when an initial gas surface density is sufficiently high (\citealt{kim2018, fukushima2020, fukushima2021}, see also \citealt{krumholz2019}). 
In fact, 
assuming the Kennicutt-Schmidt relation \citep{kennicutt1998b} and that the spatial distributions of gas and UV-emitting regions are the same,
the UV bright and compact properties of \targb\ imply a high gas density of $\Sigma_{\rm gas}\simeq1.5\times10^{4}\,  M_{\odot}$~pc$^{-2}$ or even higher out to $\simeq5.6\times10^{4}\, M_{\odot}$~pc$^{-2}$, given the current upper limit of the obscured SFR$_{\rm IR}<120\, M_{\odot}$~yr$^{-1}$ in the high-$z$ case (Section \ref{sec:continuum}). 
If we assume these gas density estimates and assume a gas-phase metallicity of $Z=0.01Z_{\odot}$, an analytical model developed in \cite{fukushima2021} suggests the star-formation efficiency to be $\simeq$ 0.7--1.0. 
Although the spec-$z$ confirmation is essentially required, the presence of \targb\ at $z \sim 17$ may not necessarily contradict with the current early galaxy evolution models and underlying $\Lambda$CDM frame work, based on the observed properties so far. 

\subsection{Dust poor universe at $z\gtrsim11$}
\label{sec:irx}

Recent ALMA observations for UV-bright galaxies dominating the bright-end of UV luminosity function (LF) show successful detection of the dust continuum from $\sim$40 \% of the sample  at $z\sim7$ \citep{bouwens2021, inami2022}. In contrast, we do not detect robust continuum detection from any of the $z\sim11$--17 candidates, although they also dominate the bright-end of the UV LF at these redshifts \citep[e.g.,][]{harikane2023, donnan2022}. 
This might imply that a transition is taking place in dust properties of early galaxies between $z\gtrsim11$ and $z\sim7$. 

In Figure \ref{fig:irx}, we show our measurements of the infrared excess IRX ($\equiv L_{\rm IR}/L_{\rm UV}$), UV continuum slope $\beta_{\rm UV}$, and $M_{\rm star}$ for the $z\sim11$-17 candidates. We evaluate the $L_{\rm IR}$ values with the single MBB based on the following two assumptions: the $T_{\rm d}-z$ relation of \cite{sommovigo2022} and a constant value of $T_{\rm d}=$ 50~K. The other measurements are taken from the best-fit results from {\sc cigale} summarized in Table \ref{tab:cigale_output}. 
For comparison, we also present the measurements obtained in other high-$z$ star-forming galaxies in recent ALMA large surveys of ASPECS at $z\sim2$--3 \citep[e.g.,][]{bouwens2020}, ALPINE at $z\sim4$--6 \citep[e.g.,][]{fudamoto2020, burgarella2022}, and REBELS at $z\sim7$ \citep[e.g.,][]{inami2022}. 
We find that the UV bright $z\sim$11--17 candidates are generally characterized as bluer and less IR-bright systems than the REBELS galaxies, despite similar $M_{\rm star}$ values. 
\cite{ziparo2022} discuss two possible scenarios for relatively massive ($M_{\rm star}\sim10^{8-9}\,M_{\odot}$) and blue ($\beta_{\rm UV}<-2.0$) high-$z$ ($z>10$) candidates identified in recent \jwst\ observations: a) ejected by the radiation pressure (see also \citealt{ferrara2022}), or b) segregated with respect to UV-emitting regions. Because the non-detection of the dust continuum disfavors the scenario b), the massive and blue properties observed in the $z\sim11$--17 candidates likely support the scenario a). 

We note that not all of the upper limits of IRX in the $z\sim$11--17 candidates are similarly deep as the lowest IRX regime observed in the ALPINE and REBELS results. Thus, there is a possibility that 
these $z\sim$11--17 candidates also follow the IRX relations similar to the $z\sim$2--7 galaxies, while the upper limits of ALMA might be still insufficient to capture the dust emission from them. 
Nevertheless, the parameter space currently constrained by HD1 already explores the most massive, bluest, and IR-faintest regimes which deviate from the relations evaluated by stacking for ASPECS and ALPINE sources at $z\sim2$--6. In addition to its very massive ($M_{\rm star}\sim10^{10}\,M_{\odot}$) aspect at $z\sim13$ in the $\Lambda$CDM framework \citep[e.g.,][]{steinhardt2016, steinhardt2022, mason2022, boylan2022, lovell2022, menci2022}, HD1 will be the most challenging object also with respect to dust properties, once the redshift is spectroscopically confirmed.

\section{Other Potential Low-\lowercase{$z$} Interlopers}
\label{sec:low-z}

Along with the discussions in \cite{zavala2022} and \cite{naidu2022b}, our initial SED analysis confirms that lower-$z$ line-emitting red objects can reproduce clear dropout features in the NIR filters, which resembles the Lyman-$\alpha$ break feature from very high-$z$ galaxies (Section \ref{sec:photoz}). Although we rule out the possibility of lower-$z$ DSFGs with SFR of $>30\, M_{\odot}$~yr$^{-1}$ for \targb\ and similar constraints obtained in other three candidates, owing to the deep constraints on dust continuum emission from our and recent ALMA observations (Section \ref{sec:sed}), caution is still required given the presence of populations other than DSFGs that might also play a part of the line-emitting red continuum objects, such as dusty quasars (QSOs) and active galactic nuclei (AGN) emerged in quiescent galaxies (QGs) (see also discussion in \citealt{naidu2022b}). 
In particular, more caution may be required when the objects are remarkably luminous and high redshift, where the abundance can be overwhelmed by rare populations at lower redshifts.
Note that all the candidates at $z\sim$11--17 studied in this paper, except for HD1 which was originally identified from ground-based telescopes and {\it Spitzer}, have been observed with spatially-resolved morphology in the superb resolution of JWST/NIRCam images \citep[e.g.,][]{ono2022, yang2022}. 
The spatially-resolved morphology suggests that these candidates are unlikely type-I QSOs with a point-source morphology. However, there still remains a possibility of type-II QSOs or very faint type-I QSOs, where the contrast of the host galaxy to the central QSO becomes high. 
Given these potential contributions from lower-$z$ rare objects, we investigate the remaining lower-$z$ possibility from three aspects: i) EW distribution of the optical emission lines, 
ii) optical--mm SED properties, and iii) abundance in the following subsections. 
Given the requirement of the red continuum and strong emission lines for the lower-$z$ interlopers to make the NIR dropout feature, we focus on the following two populations: type-II and/or dusty type-I QSOs/AGNs, and QG harboring AGN (QG+AGN).

\subsection{Distribution of EW(\oiii+H$\beta$)}
\label{sec:ew_dist}

\begin{figure}
\includegraphics[trim=0cm 0cm 0cm 0cm, clip, angle=0,width=0.48\textwidth]{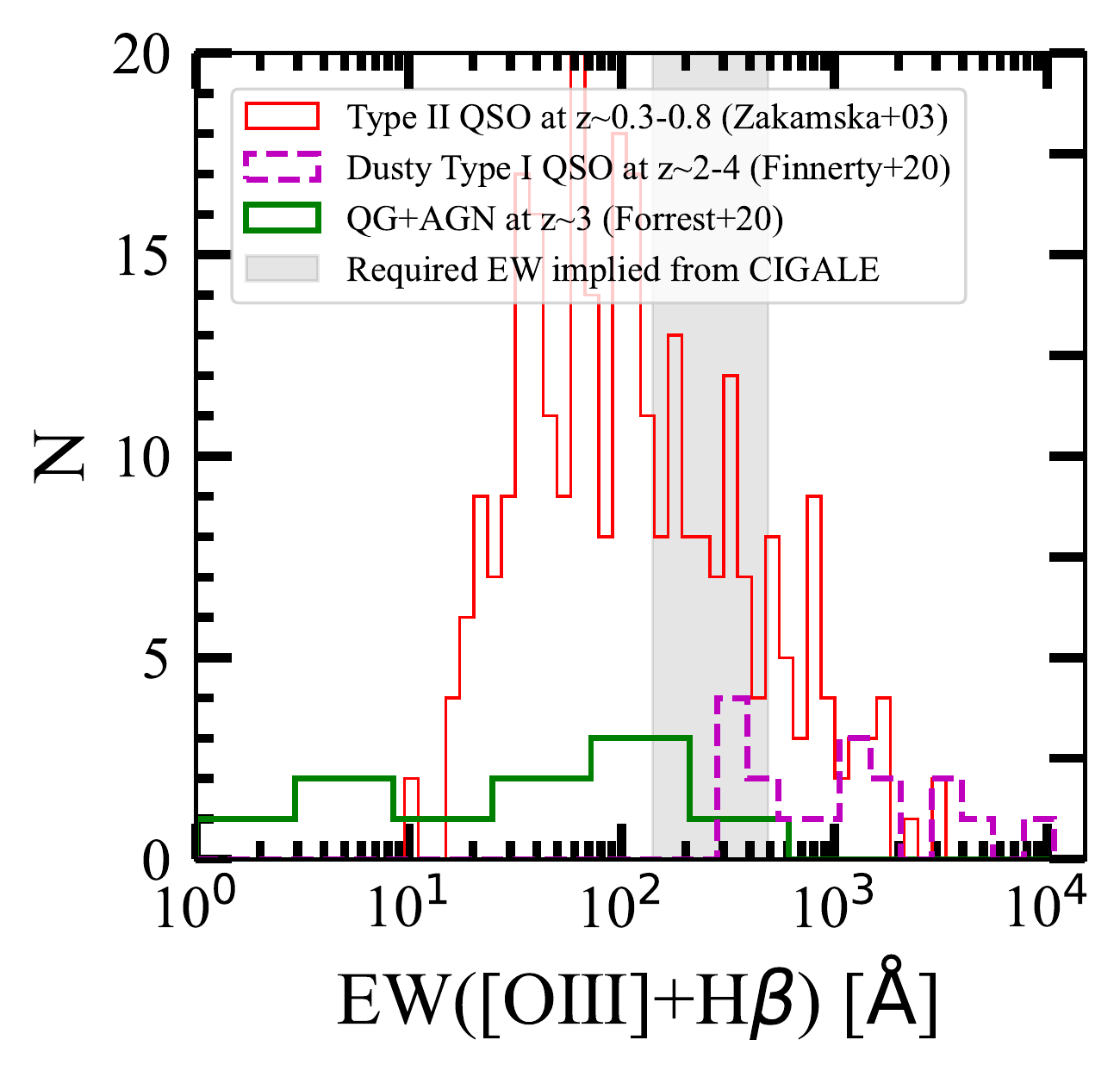}
\caption{
Distribution of the rest-frame EW(\oiii+H$\beta$) for several quasar and galaxy populations that might attribute to the $z\sim$11--17 candidates: type-II QSOs \citep{zakamska2003}, dusty-reddened type-I QSOs \citep{finnerty2020}, and QGs with emission lines from AGN \citep{forrest2020}. 
The grey shade indicates the required EW(\oiii+H$\beta$) range implied from the {\sc cigale} modeling, showing the minimum to maximum range among the best-fit SEDs of S5-z17-1, GHZ1/GLz13, GHZ2/GLz11, and HD1 forced at $z<8$. 
When the literature does not show the H$\beta$ line measurement, we include the H$\beta$ contribution by assuming the typical line ratio of \oiii/H$\beta$ from \cite{richardson2014}.
\label{fig:ew_oiii}}
\end{figure}

First, we examine the distributions of EW(\oiii+H$\beta$) for type-II / dusty type-I QSO and QG+AGN populations that might contribute to the NIR dropout objects. 
Note that the emission lines of ionized gas have been identified in QGs likely due to AGNs \citep[e.g.,][]{belli2017, belli2019, kubo2022, ito2022}. 
In the optical--NIR SED analysis forced to the lower-$z$ solution (Section \ref{sec:photoz}), we find that a dusty galaxy with EW(\oiii+H$\beta) = 450 {\rm \AA}$ reproduces the F200W dropout feature of \targb. In the same analysis for GHZ1/GLz11, GHZ/GLz13, and HD1, we obtain an EW range of EW(\oiii+H$\beta) = 140$--$490 {\rm \AA}$ from the best-fit SEDs forced at lower-$z$. 
Because more robust dropout features can be produced with higher EW values, we regard the range of EW(\oiii+H$\beta) = 140$--$490 {\rm \AA}$ as the minimum required EW values for the lower-$z$ interlopers to contaminate the high-$z$ candidates ($z\gtrsim11$) in the following analysis. 

In Figure \ref{fig:ew_oiii}, we show the distributions of EW(\oiii+H$\beta$) for type-II / dusty type-I QSO\footnote{
While the sample is called Hot Dust Obscured Galaxies (Hot DOGs) in the literature, we refer it to dust reddened type-I QSOs because of the clear detection of broad emission lines \citep{finnerty2020}.} and QG+AGN populations \citep{zakamska2003, finnerty2020, forrest2020}\footnote{
In \cite{forrest2020}, we regard 10 galaxies with $\log$(sSFR) $<-1$ Gyr$^{-1}$ and \oiii+H$\beta$ line detection as QG+AGN. 
}. 
For comparison, we also show the minimum required EW values for the lower-$z$ interlopers (grey shade). 
Based on the distribution and the lower bound of the grey shade, we find that $\sim40$\% ($\sim$100\%) of the type-II QSOs (dusty type-I QSOs) fall in and above the minimum required EW range and that the maximum EW(\oiii+H$\beta$) value reaches $\sim$3000~${\rm \AA}$ ($\sim$9000~${\rm \AA}$). 
We also find that the QG+AGN population has the EW(\oiii+H$\beta$) distribution out to $\sim$300~${\rm \AA}$, where the $\sim$40\% of them fall in and above the minimum required EW range. 
Because about 10\% of QGs at high redshift harbor emission lines are likely powered by the AGN \citep[e.g.,][]{belli2017a, belli2019}, we estimate $\sim4$\% ($= 0.1\times0.4$) of the QGs satisfy the minimum required EW range. 
By stacking Keck/NIRES spectra, an average EW of the hot obscured dusty objects at $z\sim1-4$ is also estimated to be $\sim400$~${\rm \AA}$ \citep{mackinney2023}. 
These results indicate that subsets of QSO and QG populations may actually be included in the high-$z$ ($z\gtrsim11$) candidates by contributing to the NIR dropout feature with the red continuum and the strong emission lines.

\subsection{Optical--mm SED analysis}
\label{sec:template_fit}

\begin{figure*}
\includegraphics[trim=0cm 0cm 0cm 0cm, clip, angle=0,width=0.95
\textwidth]{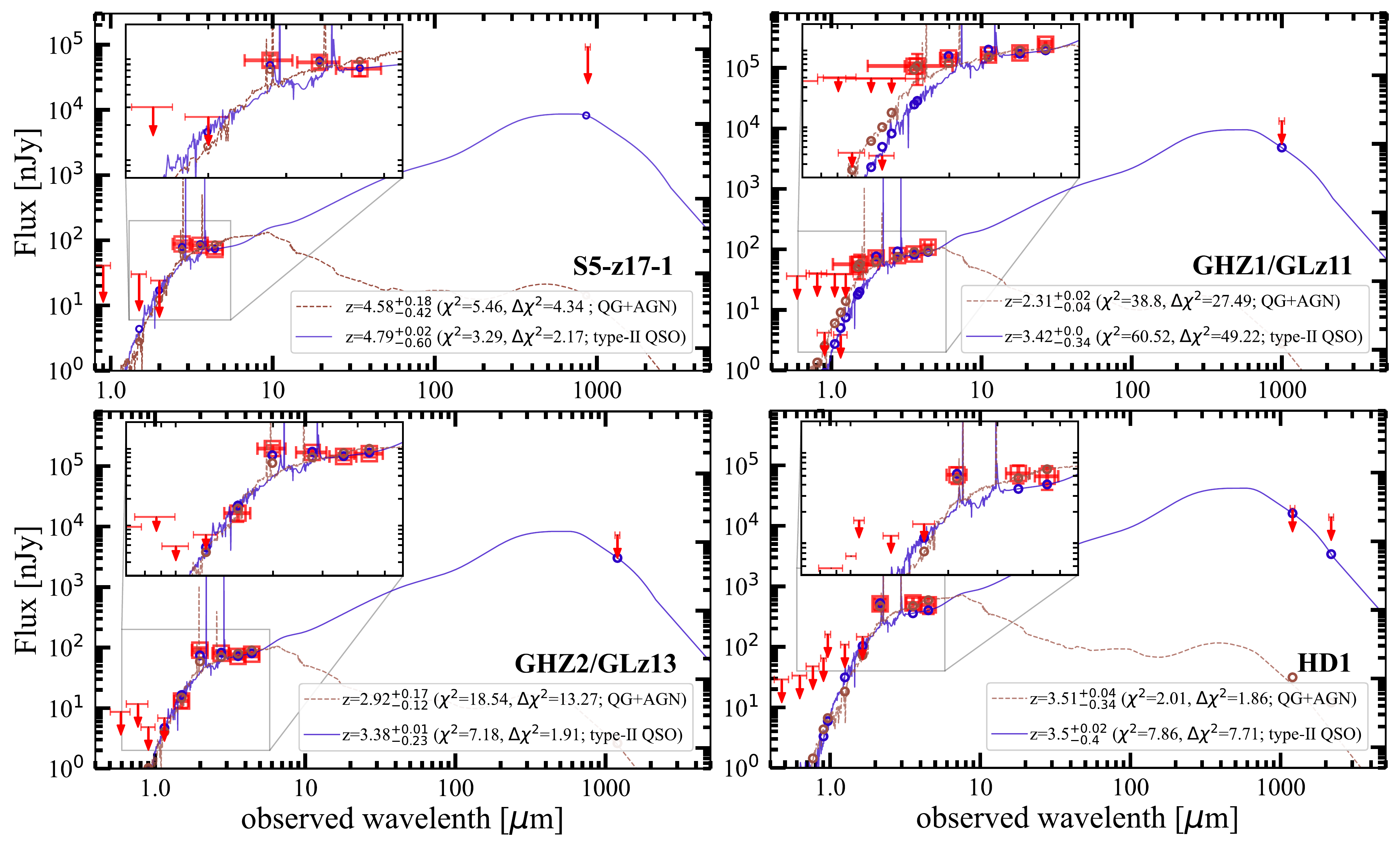}
\caption{
Same as Figure \ref{fig:nircam_alma}, but assuming QSO/AGN populations. 
The dark blue and brown dashed curves show the best-fit SED templates of type-II QSO and QG+AGN populations, respectively, from the fitting range at $0<z<20$. 
Both templates favor the lower-$z$ solution ($z\sim2$--5). 
The $\Delta\chi^{2}$ value in the label indicates the difference of the $\chi^{2}$ value from the best-fit high-$z$ galaxy solution shown in Figure \ref{fig:nircam_alma}. 
All candidates, except for GHZ1/GLz11, have the reasonable solutions at lower-$z$ with $\Delta\chi^{2}\lesssim4$. 
\label{fig:template}}
\end{figure*}

Second, we examine the optical--mm SED properties with the following two populations in this subsection: 1) type-II QSOs and 2) QG+AGN. 
Based on the EW(\oiii+H$\beta$) distribution of each population in Figure \ref{fig:ew_oiii}, we assume EW(\oiii+H$\beta$) = 1000~${\rm \AA}$ and 300~${\rm \AA}$ for the type II QSO and the QG+AGN populations, respectively, by boosting the key optical emission lines of \oiii4959, 5007, H$\beta$, H$\alpha$, and \nii\ in the type-II QSO and QG templates taken from \cite{polletta2006, polletta2007}. 
We follow the line ratios of the most highly ionized system in \cite{richardson2014}. 

In Figure \ref{fig:template}, the dark blue and brown dashed curves present the type II QSO and the QG+AGN templates fitted to the $z\sim11$--17 candidates, respectively.  
We carry out these SED template fits at $0 < z < 20$ and 
obtain the best-fit redshifts at $z\sim2$--5.    
Although the $\chi^{2}$ values are still larger than that of the best-fit high-$z$ galaxy solution with {\sc cigale} (Section \ref{sec:sed}), all candidates, except for GHZ1/GLz11, show the type-II QSO and/or QG+AGN solutions with $\Delta\chi^{2}$ values from the best-fit high-$z$ galaxy solution smaller than $\sim$4 that is lower than the criterion generally used for the high-$z$ galaxy candidate selection \citep[e.g.,][]{bowler2020, harikane2022b, donnan2022, finkelstein2023}. 
\cite{kaasinen2022} also revisit the SED fitting for HD1 with the new ALMA photometry in both 1-mm and 2-mm bands by using {\sc magphys} \citep{dacunha2015} and obtain $\chi^{2}=2.32$ from a low-$z$ solution at $z=3.98$ with a QG template. 
These results indicate the low-$z$ solutions can be plausible in some of the high-$z$ candidates even with the clear NIR dropout feature and the stringent submm--mm upper limits.

\subsection{Abundance}
\label{sec:qso_lf}

\begin{figure*}
\includegraphics[trim=0cm 0cm 0cm 0cm, clip, angle=0,width=1.0\textwidth]{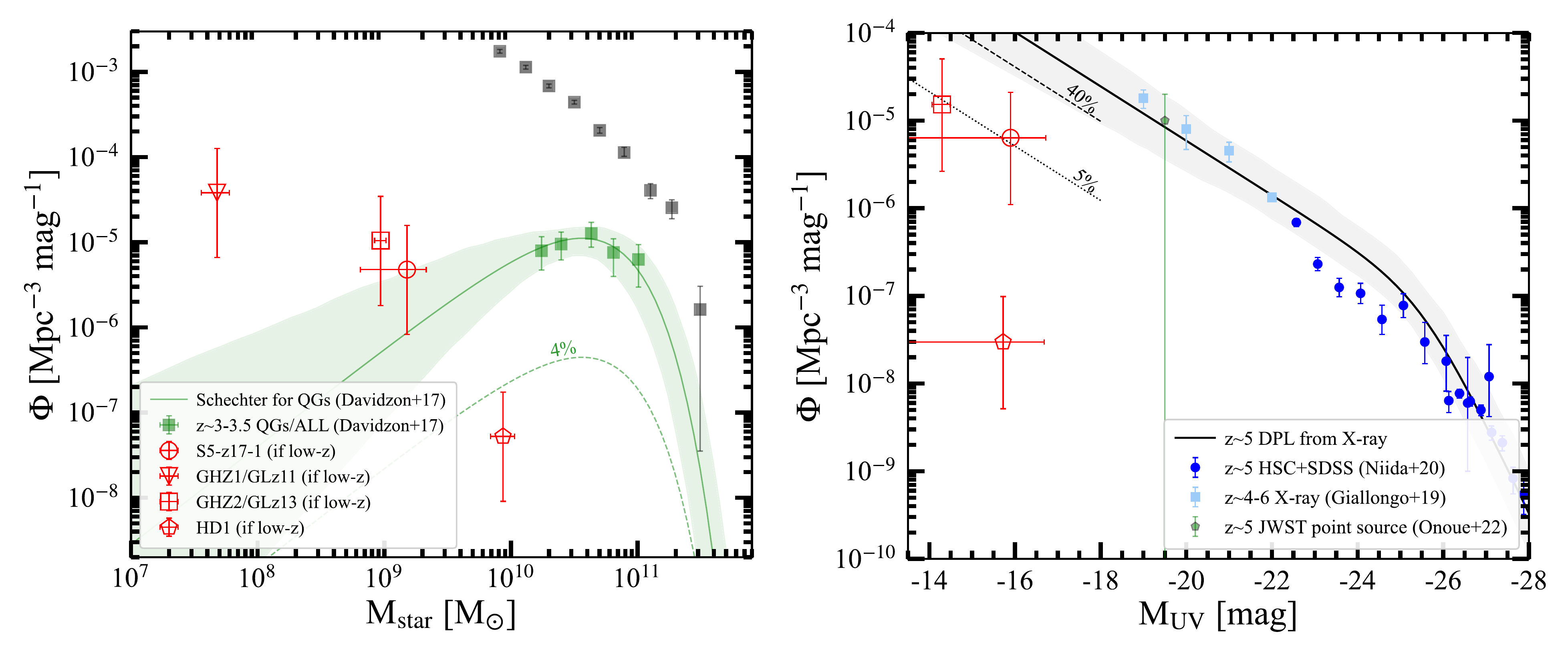}
\caption{
{\it Left:} Stellar mass function (SMF) for galaxies. The red symbols represent the abundance of the $z\sim11$--17 candidates in the case that they are the lower-$z$ line-emitting red objects at $z\sim$2--5.   
The grey and green squares show the COSMOS2015 results for entire and quiescent galaxies (QGs) at $z\sim3$--3.5 \citep{davidzon2017}. 
The green curve with the shaded region is the best-fit Schechter function with $1\sigma$ error for $z\sim$3--3.5 QGs \citep{davidzon2017}. 
{\it Right:} LF for $z\sim4$--5 QSOs/AGNs compiled from the literature. The red symbols are the same as the left panel. 
The black line with the shaded region is the best-fit Double Power Low (DPL) function with $1\sigma$ error estimated for X-ray QSOs/AGNs \citep{giallongo2019}.
In both panels, the dashed curve represents the possible abundance of the QGs and the type-II and/or dusty type-I QSOs/AGNs with strong emission lines by scaling the QG SMF and the QSO/AGN LF by 4\% and 40\%, respectively, based on their EW(\oiii+H$\beta$) distributions (Section \ref{sec:ew_dist}), indicating that these populations can be abundant low-$z$ interlopers in the survey volumes used for the identification of some of these $z\sim$11--17 candidates. 
The dotted curve in the right panel shows the 5\% scaling case for reference. 
The $M_{\rm star}$ and $M_{\rm UV}$ values are taken from the best-fit SEDs forced at $z<8$ with {\sc cigale} (green curves in Figure \ref{fig:nircam_alma}). 
GHZ1/GLz11 in this forced lower-$z$ solution shows $M_{\rm UV}\sim-10$ mag, located outside of the figure in the right panel. 
\label{fig:qso-qg_lf}}
\end{figure*}

Finally, we compare the abundance of the $z\sim$11--17 candidates with those of lower-$z$ QG and QSO/AGN populations. 
Figure \ref{fig:qso-qg_lf} presents the stellar mass function (SMF) for galaxies including QGs (left panel) and the LF for QSOs/AGNs (right panel) at $z\sim3$--5 \citep{davidzon2017, niida2020, mcgreer2018, giallongo2019, onoue2022}. 
We use the $M_{\rm star}$ and $M_{\rm UV}$ values of the $z\sim11$--17 candidates estimated from the best-fit SEDs with {\sc cigale} in the lower-$z$ case at $z\sim$2--5. 
To avoid the uncertainty of the dust attenuation correction, we use the observed-frame $M_{\rm UV}$ estimate. 
Because \targb, GHZ1/GLz11, are GHZ2/GLz13 are the most luminous high-$z$ candidates identified in the early \jwst\ data at each redshift, we conservatively adopt the survey area of 90.4~arcmin$^{2}$ from SMACSJ0723, GLASS, CEERS, and Stephan's Quintet fields \citep{harikane2023}, while we use the survey volume of 2.3~deg$^{2}$ for HD1 from \citep{harikane2022b}. 
We evaluate the possible redshift range $\Delta z$ from the 2$\sigma$ range of the $z_{\rm phot}$ estimates in the best-fit SEDs forced at $z=2$--5, resulting in $\Delta z\sim0.2$--0.8, depending on the candidate. 
We include the $1\sigma$ Poisson uncertainty presented in  \cite{gehrels1986}. 
Note that NIRCam medium-band filters are helpful to limit the possibility of the low-$z$ contamination to a very narrow redshift window of $\Delta z\lesssim0.1$ \citep{naidu2022b, arrabal-halo2023a}, while none of these four candidates have been observed with the medium-band filters, and it is not the case here.  
Another note is that the $M_{\rm UV}$ value of GHZ1/GLz11 in this forced lower-$z$ case shows $\sim-10$ mag which locates outside of the right panel, while the abundance is estimated to be $\sim3\times10^{-5}$~Mpc$^{-3}$~mag$^{-1}$ similar to other candidates. Such a very small $M_{\rm UV}$ value is required from the NIR dropout feature of GHZ1/GLz11 between F115W and F150W which is the most significant by $\sim$2.9 mag among these four candidates.

In the SMF for galaxies, 
the green curve is drawn from the best-fit Schechter function estimated for $z\sim3$--3.5 QGs \citep{davidzon2017}. 
We find that the volume densities of GHZ1/GLz11 is much higher than the abundance of QGs by more than one order of magnitude beyond the errors.  
If we take the $\sim4$\% into account as the possible fraction of the QG+AGN population that has strong enough emission lines among the QGs (Section \ref{sec:ew_dist}), the deviation becomes even significant (green dashed curve), and the abundances of GHZ2/GLz13 and \targb\ also fall above more than one order of magnitude than that of the QG+AGN population beyond the error. 
This indicates that the QG+AGN population is too rare to contaminate the $z\sim$11--17 galaxy selection in their survey volumes. 
On the hand, we find that the volume density of HD1 is much below the QG+AGN populations beyond the errors, suggesting that the QG+AGN population is an abundant contaminant in the survey volume of HD1. 
These results suggest that the possibility of contamination from the QG+AGN population is negligible in the $z\sim11$--17 candidates, except for HD1.
We note that the faint end of the QG SMF at $z\sim3.0-3.5$ could be rather flat, instead of the turnover shape\footnote{
The turnover shape is obtained at $z\sim2.5-3.0$, which is fixed in the $z>3$ measurements in \cite{davidzon2017}.}. However, the faint-end extrapolation for the QG+AGN population (green dashed curve) with such a flattened shape still falls below the volume densities of \targb, GHZ1/GLz11, and GHZ2/GLz13, and thus the above argument is unchanged. 

In the QSO/AGN LF, the black curve shows the best-fit Double Power Low (DPL) function from the X-ray QSO/AGN observations. 
Note that this is a steeper faint-end slope and a higher abundance at $M_{\rm UV}>-22$ by $\sim1$--2 orders of magnitudes than the best-fit measurement from the UV observations \citep[e.g.,][]{niida2020, finkelstein2022a}.  
While these previous measurements are still consistent within the uncertainties, the main reason would be that the X-ray observations retrieve populations such as type-II and dusty obscured QSOs/AGNs that are generally missed in the UV observations. In fact, the high fraction ($\gtrsim$80--90\%) of the obscured QSOs/AGNs at $z\gtrsim4$ have been supported from multiple aspects both from observations \citep[e.g.,][]{eilers2018, vito2018b, davies2019, morishita2020, fujimoto2022, endsley2022a} and simulations \citep[e.g.,][]{ni2020, gilli2022}. 
We thus regard that the faint end of the QSO/AGN LF from the X-ray observations is mostly dominated by the type-II and/or dusty obscured QSOs/AGNs. 
It is worth mentioning that recent \jwst/NIRSpec observations routinely identify broad-line AGNs and subsequently infer their abundance is close to the faint-end of the X-ray-based QSO/AGN LF \citep[e.g.,][]{kocevski2023, harikane2023b}. 

By extrapolating the faint end of the best-fit DPL, we find that the extrapolation exceeds the abundance of the $z\sim11$--17 candidates by more than one order of magnitude. If we take the $\sim$40\% into account as the possible fraction of the type-II QSO population that has strong enough emission lines (Section \ref{sec:ew_dist}), the abundance of the $z\sim$11--17 candidates is still far below the extrapolation (black dashed curve). 
For reference, we find that a scaling factor of $\sim5\%$ (black dotted curve) provides the comparable abundance between the $z\sim$11--17 candidates and the faint-end of the QSO/AGN LF. 
From the EW(\oiii+H$\beta$) distribution, 
the fraction of the objects with the minimum required EW(\oiii+H$\beta$) of $\gtrsim300\, {\rm \AA}$ comfortably surpasses the 5\% among the type-II and dusty QSO/AGN populations.  
This indicates that the type-II and/or dusty QSOs/AGNs with strong emission lines may overwhelm the abundance of the $z\sim11$--17 candidates in the $M_{\rm UV}$ range and indeed contaminate the $z\sim11$--17 candidates 
and that the secondary peak in $P(z)$ may not be negligible. 
For example, the middle panel of Figure~\ref{fig:nircam} suggests $P(z)$ at the secondary lower-$z$ peak at $z\sim5$ is $\sim1$--20\% in \targb, which may have the comparable probability if the abundance of the specific lower-$z$ populations exceeds that of the high-$z$ galaxies by $\sim\times5$--100.   
Although we first need to understand what lower-$z$ populations are exactly the contaminants to accurately evaluate the abundance excess of such populations, these results urge the importance of taking the high surface density of the lower-$z$ contaminants into account in the ultra high-redshift galaxy search.

Observations with an additional NIRCam medium-band filter limit the possibility of low-$z$ contamination to a very narrow redshift window ($\Delta z\lesssim0.1$; e.g., \citealt{naidu2022b}). This strategy helps to mitigate the probability of low-$z$ contaminants. However, it is worth noting that another $z\sim17$ candidate, \targa, despite also being observed with the medium-band filter of F410M, has been spectroscopically confirmed at $z=4.91$ \citep{arrabal-halo2023a}. 
This also remarks on the high surface density of the lower-$z$ contaminants.  

\subsection{Remaining low-$z$ possibilities}
\label{sec:remaining}

\setlength{\tabcolsep}{8pt}
\renewcommand{\arraystretch}{1.5}
\begin{table*}
\begin{center}
\caption{Remaining low-$z$ possibilities from $\Delta\chi^{2}$ and abundance perspectives}
\label{tab:lowz}
\vspace{-0.4cm}
\begin{tabular}{ccccccc}
\hline \hline
Source Name & Possible low-$z$ population & $z_{\rm phot}$ ($\chi^{2}$) & $\Delta {\chi^{2}}^{\dagger}$  & Note  \\ \hline 
\multirow{2}{*}{S5-z17-1} &  type-II or dusty type-I QSO/AGN & $4.79^{+0.02}_{-0.60}$ (3.29) & 2.17 &  
Very compact ($r_{\rm e}=0\farcs05^{+0\farcs03}_{-0\farcs02}$)   \\
&  (QG+AGN) & $4.58^{+0.18}_{-0.42}$ (5.46) & 4.34 &    Unlikely from the possible abundance \\
\hline
\multirow{2}{*}{GHZ1/GLz11} & \multirow{2}{*}{--}  & \multirow{2}{*}{--}  &  \multirow{2}{*}{--} & The most secure candidate at $z\gtrsim11$ \\  
 &   &  &  & owing to [F115W]$-$[F150W] $>$ 2.9 mag \\  \hline
GHZ2/GLz13 & type-II or dusty type-I QSO/AGN & $3.38^{+0.01}_{-0.23}$ (7.18) & 1.91 & Very compact ($r_{\rm e}=0\farcs02^{+0\farcs01}_{-0\farcs00}$) \\ \hline
HD1 & QG+AGN & $3.51^{+0.04}_{-0.33}$ (2.01)  & 1.86 &   \\ \hline \hline
\end{tabular}
\end{center}
$\dagger$ Difference of the $\chi^{2}$ value from the best-fit high-$z$ galaxy solution at $z\sim11$--17 summarized in Table \ref{tab:cigale_output}, 
suggesting that high-$z$ solution is still favored in every candidate. 
\end{table*}
\renewcommand{\arraystretch}{1}

In Section \ref{sec:ew_dist} and \ref{sec:template_fit}, we find that the subsets of QG and QSO/AGN populations actually have strong enough optical emission lines which produce the reasonable SED fits ($\Delta\chi^{2}\lesssim4$) in some of the $z\sim11$--17 candidates. 
In Section \ref{sec:qso_lf}, we confirm that the abundance of such type-II and/or dusty type-I QSOs/AGNs with strong enough emission lines is higher than that of the $z\sim11$--17 candidates, while the abundance of such QG populations is negligible, except for HD1. 
These results indicate the need to consider the relative surface densities of lower-$z$ contaminants in the ultra-high-$z$ galaxy search. 

In Table \ref{tab:lowz}, we summarize the remaining low-$z$ possibilities for each candidate. 
If the abundance of the lower-$z$ population is comparable or overwhelming (see Figure~\ref{fig:qso-qg_lf}) and the low-$z$ solution shows $\Delta\chi^{2}\lesssim4$ from the best-fit high-$z$ solution (see Figure~\ref{fig:template}), we regard the low-$z$ solution as the remaining possibility. 
This makes the QG+AGN solution in \targb\ unlikely plausible because of its negligibly small abundance (Section \ref{sec:qso_lf}).
We find that GHZ1/GLz11 denies all lower-$z$ solutions, showing $\Delta\chi^{2} > 20$ in every type of the lower-$z$ object we investigate in this paper. The reason is simply because of the fact that the most robust dropout feature is observed in GHZ1/GLz11 between the F115W and F150W filters by $\sim2.9$ mag (cf. $\sim$1.6--2.1~mag in the other three candidates). 
On the other hand, the other three sources all have the remaining low-$z$ solutions both from $\Delta\chi^{2}$ and abundance perspectives, indicating that the low-$z$ solutions cannot be ruled out in the majority of the ultra-high-$z$ galaxy candidates.

Interestingly, we find that the possibility of the type-II or dusty type-I QSOs/AGNs remains in \targb\ and GHZ2/GLz13 that fill the most UV luminous and compact parameter space among the recent \jwst\ high-$z$ candidates at $z>9$ with $r_{\rm e}=0\farcs02$--$0\farcs05$ (see e.g., Figure 18 in \citealt{ono2022}). 
While numerical simulations confirm the presence of such a compact galaxy forms at $z>10$ (\citealt{yajima2022}; see also discussion in \citealt{ono2022}), the remarkably compact size might be caused by non-negligible contribution of the emission from the QSO/AGN. This implies a very intriguing scenario of the emergence of the QSO/AGN at $z>10$, or the lower-$z$ interloper of the type-II and/or dusty type-I QSO/AGN. 
We also refer the reader to the discovery of a remarkably UV bright ($M_{\rm UV}\simeq-24.4$), compact, very blue ($\beta_{\rm UV}\simeq-2.2$), dust- and metal-poor starburst galaxy at $z=2.5$ \citep{marques2020}, which suggests that we may be witnessing similar objects at $z\gtrsim11$. Nevertheless, the rest-UV effective radius of the $z=2.5$ object is measured to be $r_{\rm e}\simeq1.2$~kpc \citep{marques2020}. 
These results suggest that \targb\ and GHZ2/GLz13 are almost 10 times more compact than the $z=2.5$ object, while the complex NIRCam PSF is not yet fully characterized and some relevant uncertainties may remain\footnote{
The difference in the PSF shape has been reported between the \jwst\ software tool \texttt{webbpsf} and the empirical approach using stars observed in the NIRCam FoV \citep[e.g.,][]{ono2022, ding2022}}.. 
Following the recent successful spectroscopic confirmations of galaxies at $z\gtrsim9$ with \jwst/NIRSpec \citep[e.g.,][]{roberts-borsani2022b, williams2022, bunker2023, curtis-lake2022, tang2023, fujimoto2023, arrabal-halo2023a, arrabal-halo2023b, hsiao2023}, confirmation of the FIR line candidate with ALMA, and/or making spectroscopic follow-up with \jwst/NIRSpec, will be crucial for these UV bright $z\sim$11--17 candidates to reach a definitive conclusion.

\section{Summary}
\label{sec:summary}
In this paper, we present the ALMA Band 7 observations of a remarkably bright and high-redshift galaxy candidate S5-z17-1 ($M_{\rm UV}=-21.6$ at $z_{\rm phot}\sim17$) with a robust NIRCam/F200W dropout feature identified in \jwst\ ERO data of Stephan's Quintet. The number of UV-bright high-$z$ candidates at $z>9$ exceeds most pre-\jwst\ predictions,  remarking the importance of testing lower-$z$ contaminants, especially from populations with a red continuum and strong emission lines which can produce similar dropout features of high-$z$ galaxies in the NIRCam filters. 
In conjunction with the other three UV-bright $z\gtrsim11$ candidates recently observed ALMA, we systematically conduct the spectral energy distribution (SED) analysis over the optical--mm wavelengths and discuss their physical properties in their high-$z$ solutions and remaining low-$z$ possibilities for each candidate. 
This is the first ALMA FIR census for the best candidates of remarkably UV-bright and high-redshift candidates at $z\gtrsim11$ from the community, including the initial FIR characterization of the F200W dropout population newly identified with \jwst. 
The main findings of this paper are summarized as follows:

\begin{enumerate}
\item Based on the SED analysis with the latest NIRCam photometry using {\sc cigale} and {\sc eazy}, we confirm that a very high-$z$ solution of $z\geq16$ is favored in \targb, while we also confirm that a red object at $z\sim4.6$ with strong emission lines with the rest-frame equivalent width of EW(\oiii+H$\beta$) = 450~${\rm \AA}$ produces the dropout feature between F200W and F277W filter.  
For plausible estimates of the surface densities of such lower-z populations, the probability of the $z\sim4.6$ solution is comparable to the high-$z$ solution, indicating that this source may lie at lower redshifts than originally claimed. 
\item We do not detect dust continuum at 866~$\mu$m from \targb, placing the 2$\sigma$ upper limit at 90.0 $\mu$Jy. We adopt the spectral dust index of 2.0 and the dust temperature of $T_{\rm d}=$ 90~K by extrapolating the $T_{\rm d}-z$ evolution model \citep{sommovigo2022} to $z=18$, which is consistent of the lower limit of $T_{\rm d}>$ 80~K obtained from the radiative equilibrium model  \citep{inoue2020, fudamoto2022} based on a clumpy ISM assumption and a very compact effective radius of $r_{\rm e}\sim140$~pc measured in \cite{ono2022}. By assuming the single modified black body, we estimate the upper limit of the infrared luminosity of $L_{\rm IR}< 1.2\times10^{12}\,L_{\odot}$ which corresponds to the star-formation rate of SFR $<120\, M_{\odot}$~yr$^{-1}$.
In the case that \targb\ is a lower-$z$ object at $z\sim4.6$, we infer $L_{\rm IR}<2.8\times10^{11}\,L_{\odot}$ and SFR $<28\,M_{\odot}$~yr$^{-1}$.
\item We identify a line feature with the 5.1$\sigma$ level at $338.726\pm0.007$~GHz exactly at the source position. By running the blind line search algorithm of {\sc Findclump}, the fidelity is estimated to be $\sim$50\% in the entire data cube, suggesting that the realistic fidelity at the source position is much high.  
We estimate the line width of FWHM $= 118\pm20$~km~s$^{-1}$ and the line intensity of $I_{\rm line}=0.35\pm0.07$ Jy~km~s$^{-1}$. Based on potential redshift solutions, this line candidate is most likely either \cii\ 158$\mu$m at $z=4.6108\pm0.0001$ or \oiii\ 52$\mu$m at $z=16.0089 \pm 0.0004$. Although systematic uncertainties remain in applications of empirical relations, we confirm that the SFR value inferred from the line luminosity is consistent with the one estimated from the upper limit of $L_{\rm IR}$ in both cases. 
Either the \jwst/NIRSpec and/or the ALMA 88~$\mu$m line follow-up will give a definitive conclusion as to which redshift solution is true. 
\item Together with three similarly UV-bright high-redshift candidates at $z\gtrsim11$ recently observed ALMA -- GHZ1/GLz11 \citep{yoon2022}, GHZ2/GLz13 \citep{bakx2022}, and HD1 \citep{harikane2022b, kaasinen2022}, we conduct the optical--mm SED analysis including the new ALMA photometry. Owing to the deep constraints from ALMA, we find that the high-$z$ solution is strengthened in every candidate as a result of very blue (UV continuum slope of $\beta_{\rm UV}\approx-2.3$) and luminous ($M_{\rm UV}\approx[-24:-21]$) system.  
\item Based on the best-fit SEDs at $z\gtrsim11$, we compare IRX ($\equiv L_{\rm IR}/L_{\rm UV}$), $\beta_{\rm UV}$, and $M_{\rm star}$ properties of these four candidates at $z\gtrsim11$ with other high-$z$ star-forming galaxies from recent ALMA studies, including the REBELS sources at $z\sim7$ \citep{bouwens2021,inami2022}. We find that the $z\gtrsim11$ candidates have generally bluer and less IR-bright properties compared to the REBELS sample, although they place a similar $M_{\rm star}$ regime and are both dominating the bright-end of the UV luminosity function at these redshifts. 
This might indicate a transition taking place in the dust properties of early galaxies between $z\gtrsim11$ and $z\sim7$ such as the powerful dust ejection due to the radiation pressure in the very early system at $z\gtrsim11$. 
We also find that HD1 explores the most massive, bluest, and IR-faintest parameter space among these high-$z$ star-forming galaxies. 
\item We also examine remaining low-$z$ possibilities due to line-emitting red objects other than dusty star-forming galaxies. We verify type-II and/or dusty type-I quasars(QSOs)/AGNs and AGNs emerged in quiescent galaxies (QGs) based on their EW(\oiii+H$\beta$) distributions, optical--mm SED properties, and their possible abundances. Given the survey volumes used for these $z\sim11$--17 candidates, we find that the abundance of the QG+AGN population is negligibly small, except for HD1, while the abundance of the type-II and/or dusty type-I QSOs/AGNs actually overwhelms all of these candidates. We also find that the SED template of the type-II QSOs and QGs including strong emission lines produces reasonable SED fits with $\Delta\chi^{2}\lesssim4$ in all candidates, except for GHZ1/GLz11 because of the most robust continuum break by $\sim2.9$~mag between F115W and F150W filters. 
These results suggest that lower-$z$ possibilities are not ruled out in several of the $z\gtrsim11$ candidates and the importance of considering the relative surface density of the lower-$z$ contaminants in the ultra-high-$z$ galaxy search.
The detailed physical process of the dust attenuation and the ionizing background associated with the QSOs/AGNs to produce the strong emission lines with the red continuum in these potential lower-$z$ interlopers are beyond this paper, while these topics need to be discussed elsewhere. 
\end{enumerate}

We thank the anonymous referee for constructive comments and suggestions with careful review.
We are grateful to Yuichi Harikane for sharing the latest photometry of \targb, helping with the comparison of the photometry, and providing useful comments on the paper.
We thank Rohan Naidu, Paola Santini, Marco Castellano, Masafusa Onoue, Andrea Ferrara, Laura Sommovigo, Vasily Kokorev, Fengwu Sun, Romain Meyer, Leindert Boogaard, and Rui Marques-Chaves for helpful comments and discussions.
This paper makes use of the ALMA data: ADS/JAO. ALMA \#2021.A.00031.S. 
ALMA is a partnership of the ESO (representing its member states), 
NSF (USA) and NINS (Japan), together with NRC (Canada), MOST and ASIAA (Taiwan), and KASI (Republic of Korea),
in cooperation with the Republic of Chile. 
The Joint ALMA Observatory is operated by the ESO, AUI/NRAO, and NAOJ. 
This work is based on observations and archival data made with the {\it Spitzer Space Telescope}, which is operated by the Jet Propulsion
Laboratory, California Institute of Technology, under a contract with NASA along with archival data from the NASA/ESA 
{\it Hubble Space Telescope}. This research also made use of the NASA/IPAC Infrared Science Archive (IRSA), 
which is operated by the Jet Propulsion Laboratory, California Institute of Technology, under contract with the National Aeronautics and Space Administration. 
The Early Release Observations
and associated materials were developed, executed, and compiled by the ERO production team:  Hannah Braun, Claire Blome, Matthew Brown, Margaret Carruthers, Dan Coe, Joseph DePasquale, Nestor Espinoza, Macarena Garcia Marin, Karl Gordon, Alaina Henry, Leah Hustak, Andi James, Ann Jenkins, Anton Koekemoer, Stephanie LaMassa, David Law, Alexandra Lockwood, Amaya Moro-Martin, Susan Mullally, Alyssa Pagan, Dani Player, Klaus Pontoppidan, Charles Proffitt, Christine Pulliam, Leah Ramsay, Swara Ravindranath, Neill Reid, Massimo Robberto, Elena Sabbi, Leonardo Ubeda. The EROs were also made possible by the foundational efforts and support from the JWST instruments, STScI planning and scheduling, and Data Management teams.
This project has received funding from the European Union’s Horizon 2020 research and innovation program under the Marie Sklodowska-Curie grant agreement No. 847523 ‘INTERACTIONS’ and from NASA through the NASA Hubble Fellowship grant HST-HF2-51505.001-A awarded by the Space Telescope Science Institute, which is operated by the Association of Universities for Research in Astronomy, Incorporated, under NASA contract NAS5-26555.
S.F. acknowledges support from the European Research Council (ERC) Consolidator Grant funding scheme (project ConTExt, grant No. 648179). 
The Cosmic Dawn Center is funded by the Danish National Research Foundation under grant No. 140.\\

Some/all of the data presented in this paper were obtained from the Mikulski Archive for Space Telescopes (MAST) at the Space Telescope Science Institute. 
The specific observations analyzed can be accessed via \dataset[10.17909/6503-5145]{https://doi.org/10.17909/6503-5145}. 
The reduced data is also available via \url{https://s3.amazonaws.com/grizli-v2/JwstMosaics/v4/index.html}.

\software{{\sc casa} (v6.2.1 \& v6.4.1; \citealt{casa2022}), \texttt{grizli} \citep{brammer2022}, 
\texttt{eazy}, 
\citep{brammer2008}}, 
\texttt{Interferopy}, \citep{boogaard2021}, 
and \texttt{CIGALE} \citep{boquien2019}.

\bibliographystyle{apj}
\bibliography{apj-jour,reference}

\appendix

\section{\jwst\ and \hst\ photmetry}
In Table~\ref{tab:nircam}, we summarize the photometry used in our SED analysis. 
The photometry is evaluated with a circular aperture in $0\farcs5$ diameter and corrected to the total flux. A potential systematic uncertainty is added by 10\% of the total flux in the error. For HD1, we use the optical--NIR photometry estimated in \cite{harikane2022b}. 

\setlength{\tabcolsep}{1pt}
\begin{table*}
\caption{\jwst\ and \hst\ photometry used in our SED analysis for $z\sim11$--17 candidates}
\vspace{-0.6cm}
\label{tab:nircam}
\footnotesize
\begin{center}
\begin{tabular}{cccccccccccccccc}
\hline 
\hline
ID & F606W & F775W & F814W & F090W & F105W & F115W & F125W & F150W & F160W & F200W & F277W & F356W & F444W \\
     &  (nJy)   &  (nJy)      &  (nJy) &  (nJy)  &  (nJy)  &  (nJy)     &  (nJy)     &  (nJy)    &  (nJy)    &  (nJy)   &  (nJy)   &  (nJy)   &  (nJy)    \\ \hline 
S5-z17-1 & \nodata &\nodata &\nodata & 8.9$\pm$20.4& \nodata &\nodata &\nodata & 10.9$\pm$13.2& \nodata & 2.4$\pm$10.5&  89.5$\pm$11.2&  84.1$\pm$10.8&  72.4$\pm$11.4\\
GHZ1 & -40.6$\pm$18.7& \nodata & -7.3$\pm$20.6&  2.7$\pm$2.1&  7.2$\pm$19.9&  -1.2$\pm$1.9&  18.6$\pm$19.5&  56.6$\pm$6.2&  57.1$\pm$24.9&  72.5$\pm$7.6&  79.1$\pm$8.2&  83.9$\pm$8.6&  108.8$\pm$11.0\\
GHZ2 & 6.1$\pm$4.5&  1.2$\pm$6.0& \nodata & -0.6$\pm$2.5& \nodata & 4.9$\pm$3.5& \nodata & 13.1$\pm$3.0& \nodata & 91.0$\pm$9.5&  80.9$\pm$8.5&  71.5$\pm$7.5&  77.7$\pm$8.0\\
\hline \hline
\end{tabular}
\end{center}
\vspace{-0.4cm}
\end{table*}

\section{Possibility of CO(3-2)}
In Section \ref{sec:scan}, we detect the FIR line at $338.726\pm0.007$~GHz at the 5.0$\sigma$ level. Apart from the \cii\ 158~$\mu$m at $z=4.6$ and \oiii\ 52$\mu$m at $z=16.0$ discussed in Section \ref{sec:line}, another possibility could be CO(3-2) at $z=0.0208\pm0.002$, because the galaxies composed of Stephan's Quintet take the range of $z=0.0193$--0.0225.\footnote{\url{http://ned.ipac.caltech.edu/}} 
Moreover, recent NIRCam observations have identified dusty star clusters in the local galaxy of VV114, where several of them show very red in F150W--F200W, but blue in F200W--F356W \citep{linden2023}. 
This implies that some specific SED shapes of the dusty star clusters might also reproduce the Lyman-$alpha$ break. 
We thus also explore the CO(3-2) possibility by verifying if the SED shape of the dusty stellar clump satisfies the NIRCam color properties of \targb. 

By using the dust-corrected SED of the star clusters in the local galaxy presented in \cite{fernandez-ontiveros2009}, we apply the dust extinction curves ($A_{\rm V} = 1, 5, 10$, and 20) of \cite{calzetti2000} to the SED and examine the SED shape at $\sim1$--5$\mu$m wavelengths. We find that the SED shape similar to the Lyman-$\alpha$ break indeed appears due to the combination of the intrinsic stellar SED shape with a peak at $\sim$1.6$\mu$m and the less dust extinction at longer wavelengths, but the break occurs only at $\sim$1--1.5$\mu$m, and the dropout feature between F200W and F277W cannot be reproduced. We thus conclude that the NIRCam color properties of \targb\ are hard to be reproduced by the local star clusters and thus the interpretation of CO(3-2) is unlikely.

\section{CIGALE parameters for the final fit}

In Table~\ref{tab:cigale_param}, we summarize the parameters and their boundaries used for the SED fitting with {\sc cigale} in Section \ref{sec:analysis}. 

\begin{table*}[htp]
\begin{center}
\resizebox{0.8\linewidth}{\height}{
\begin{tabular}{|>{\centering}p{8.0cm}|>{\centering\arraybackslash}p{3.5cm}|>{\centering\arraybackslash}p{3.5cm}|}
  \hline\hline
  {\bf Parameters} & {\bf Symbol} & {\bf Range} \\
  \hline\hline
\multicolumn{3}{c}{}\\
\multicolumn{3}{c}{\bf Delayed SFH and recent burst}\\
  \hline
 e-folding time scale of the delayed SFH & $\tau_{main}$ [Myr] & 100, 250, 500, 1000 \\
  \hline
 Age of the main population & Age$_{main}$[Myr]  & 51 log values in [1: 3.3] \\
  \hline
 Burst & f$_{burst}$  &  No burst  \\
  \hline
    \multicolumn{3}{c}{}\\
    \multicolumn{3}{c}{\bf SSP}\\
  \hline
  SSP &   & BC03 \\
  \hline
  Initial mass function &  IMF & Chabrier \\
  \hline
  Metallicity     & Z &  0.0004, 0.004, 0.02\\ \hline
    \multicolumn{3}{c}{}\\
    \multicolumn{3}{c}{\bf Nebular emission}\\
  \hline
  Ionization parameter &  logU    & -2.0 \\ \hline
  Line width [km/s]    &     ---    &  150 \\ \hline
  Gas-phase Metallicity &  zgas   & 0.0004, 0.004, 0.02 \\ \hline
  Electron density      & ne      & 100 \\
  \hline
    \multicolumn{3}{c}{}\\
    \multicolumn{3}{c}{\bf Dust attenuation law}\\
  \hline
  Color excess for both the old and young stellar populations &  E\_BV\_lines & 21 log values in [$-3$: 1.3] \\
  \hline
  Reduction factor to apply on E\_BV\_lines to compute E(B-V)s the stellar continuum attenuation & E\_BV\_factor & 1.0 \\ 
  \hline
  Bump amplitude &  uv\_bump\_amplitude &  0.0 \\
  \hline
  Power law slope & power law\_slope &  0.0 \\
  \hline
  Extinction law to use for attenuating the emission lines flux &  Ext\_law\_emission\_lines & SMC \\
  \hline
  Ratio of total to selective extinction, A\_V / E(B-V) & Rv & 3.1 \\
  \hline
    \multicolumn{3}{c}{}\\
    \multicolumn{3}{c}{\bf Dust emission (DL2014)}\\
  \hline
  Mass fraction of PAH & $q_{PAH}$ &  0.47  \\
  \hline
  Minimum radiation field &  U$_{min}$ & 5.0  \\
  \hline
  Power law slope dU/dM $\approx$ U$^\alpha$ &---  $\alpha$ & 2.0 \\
  \hline
   Dust fraction in PDRs  & $\gamma$ & 0.1  \\
     \hline
    \hline
\multicolumn{3}{c}{}\\
\multicolumn{3}{c}{\bf No AGN emission}
\\ \hline \hline
\end{tabular}
}
  \caption{{\sc cigale} modules and input parameters used for all the fits. BC03 indicates \cite{bruzual2003}, and the Chabrier IMF refers to \cite{chabrier2003}.}
  \label{tab:cigale_param}
\end{center}
\end{table*}

\begin{figure}
\includegraphics[trim=0cm 0cm 0cm 0cm, clip, angle=0,width=0.5\textwidth]{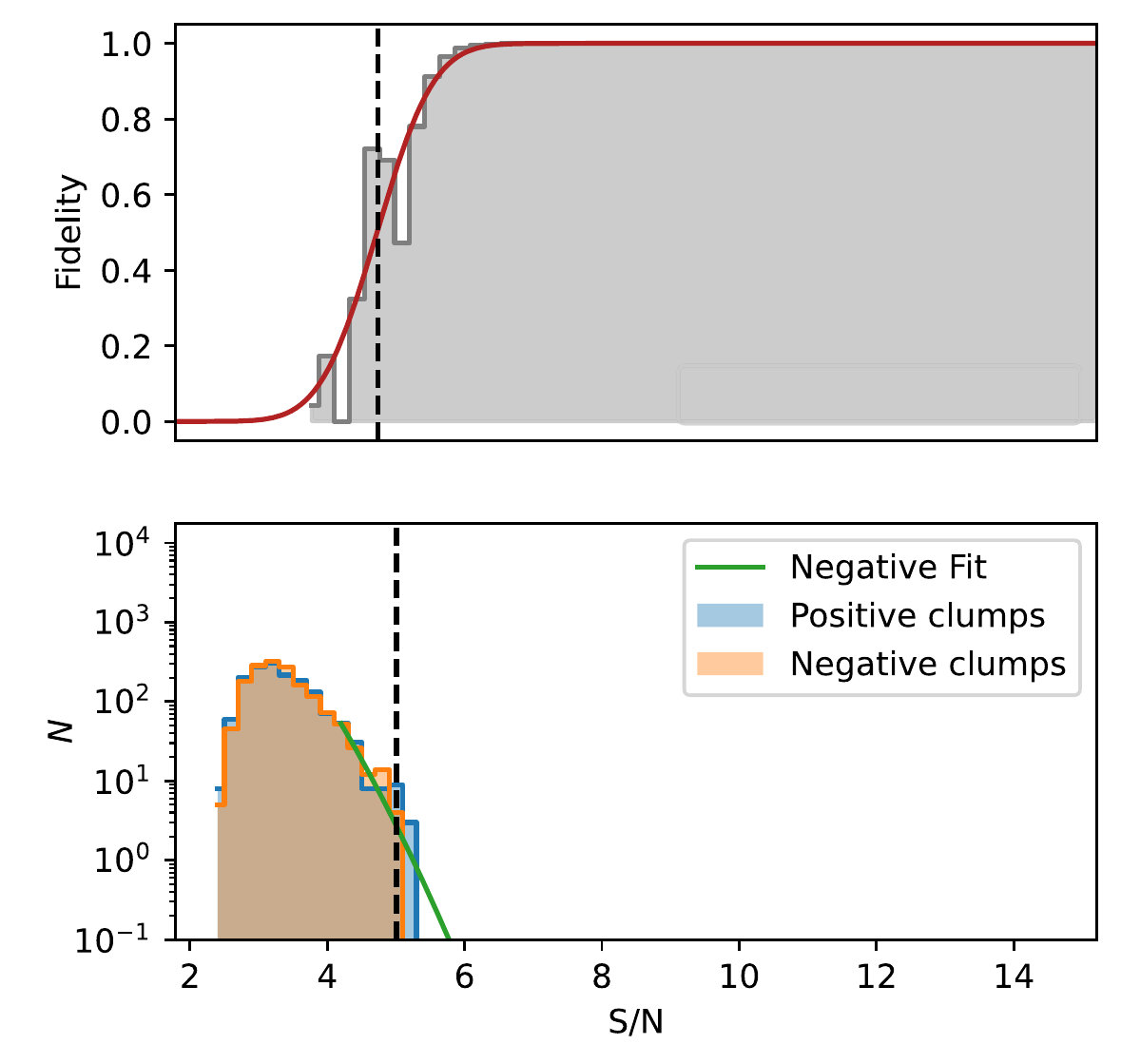}
\caption{Fidelity of the 3D clumps as a function of S/N produced by a blind line search algorithm {\sc Findclump} in our ALMA Band 7 data cube for \targb. 
Here we show the results with the 15-km width data cube smoothed with the 6-channel kernel, where the line candidate is identified with {\sc Findclump} at the 5$\sigma$ level. 
{\it Bottom:} Histograms of positive and negative clumps. {\it Top:} Fidelity estimated from the histograms of positive and negative clumps as a function of S/N. The dashed line corresponds to the $\sim$338.7-GHz line feature exactly at the source position, indicating that the fidelity is $\sim$50\%. 
Note that this is a blind search in the entire cube. Based on the survey volume only around the central target, the realistic fidelity should be much higher than 50\%.  
\label{fig:line_search}}
\end{figure}

\end{document}